\documentclass[floatfix,aps,prb,amsmath,amssymb,reprint,titlepage,twocolumn]{revtex4-2}

\usepackage{float}
\usepackage{graphicx}% Include Fig. files
\usepackage{dcolumn}% Align table columns on decimal point
\usepackage{bm}% bold math
\usepackage{lipsum}
\usepackage{titlesec}
\usepackage{epstopdf}
\usepackage{multirow}
\usepackage{siunitx}
\usepackage{color}
\usepackage{verbatim}
\usepackage[english]{babel}
\usepackage[T1]{fontenc}
\usepackage[utf8]{inputenc}
\usepackage[mathlines]{lineno}% Enable numbering of text and display math
\usepackage{afterpage}
\usepackage{soul}

\addto\captionsenglish{}
\newcommand{\bextz}{\ensuremath{B_{\text{ext},z}}}
\newcommand{\Prim}{\ensuremath{P_{\text{rim}}}}
\newcommand{\Arim}{\ensuremath{A_{\text{rim}}}}
\newcommand{\Aunit}{\ensuremath{A_{\text{unit}}}}

\usepackage[colorinlistoftodos]{todonotes}

\begin{document}
\title{Exploration of magnon-magnon coupling in an antidot lattice: \\ The role of non-uniform magnetization texture
}

\author{Mathieu Moalic}
\email{matmoa@amu.edu.pl}
\affiliation{Institute of Spintronics and Quantum Information, Faculty of Physics, Adam Mickiewicz University, Poznan, Poland}
\author{Mateusz Zelent}
\affiliation{Institute of Spintronics and Quantum Information, Faculty of Physics, Adam Mickiewicz University, Poznan, Poland}
\author{Krzysztof Szulc}
\affiliation{Institute of Spintronics and Quantum Information, Faculty of Physics, Adam Mickiewicz University, Poznan, Poland}
\author{Maciej Krawczyk}
\affiliation{Institute of Spintronics and Quantum Information, Faculty of Physics, Adam Mickiewicz University, Poznan, Poland}

\date{\today}

\begin{abstract}
We numerically study the spin wave dynamics in an antidot lattice based on a Co/Pd multilayer structure with reduced perpendicular magnetic anisotropy at the edges of the antidots. This structure forms a  magnonic crystal with a periodic antidot pattern and a periodic magnetization configuration consisting of out-of-plane magnetized bulk and in-plane magnetized rims.  Our results show the different behavior of spin waves in the bulk and in the rims under varying out-of-plane external magnetic field strength, revealing complex spin-wave spectra and hybridizations between the modes of these two subsystems. A particularly strong magnon-magnon coupling, due to exchange interactions, is found between the fundamental bulk spin-wave mode and the second-order radial rim modes. However, the dynamical coupling between the spin-wave modes at low frequencies, involving the first-order radial rim modes, is masked by the changes in the static magnetization at the bulk-rim interface with magnetic field changes. The study expands the horizons of magnonic-crystal research by combining periodic structural patterning and non-collinear magnetization texture to achieve strong magnon-magnon coupling, highlighting the significant role of exchange interactions in the coupling.
\end{abstract}

\begin{titlepage}
\maketitle
\end{titlepage}

\section{Introduction}
Magnonic crystals (MCs) are a type of magnetic meta-material characterized by the periodicity of some material properties that cause the formation of an artificially tailored spin-wave (SW) band structure.~\cite{Gulyaev2001,Puszkarski03,Lenk2011,Krawczyk14,Nikitov2015,Rychly2015,Chumak2017}. These crystals are analogous to photonic crystals, which utilize periodic modulation of the refractive index to control the propagation of electromagnetic waves~\cite{Joannopoulos08}. MCs provide guidance and control over SWs, promising them for a variety of applications~\cite{Krawczyk14,Chumak2017}. They can be created by the periodic arrangement of two different materials, which can be two ferromagnetic materials~\cite{Tacchi12}, ferromagnetic and non-magnetic, it is an array of ferromagnetic dots in a non-magnetic matrix~\cite{Kruglyak2010,Tacchi2011,Saha2013}, or inverse structures, i.e., an array of holes in a ferromagnetic matrix, also known as antidot lattices (ADLs)~\cite{Kostylev2008,Mandal2013}.

Periodic modulation of SW-relevant parameters can be introduced during the patterning process~\cite{Martin2003,Lau2011}, but also through post-fabrication techniques such as ion irradiation, which selectively modifies the magnetic properties of targeted regions in ferromagnetic films or multilayers~\cite{Carter1982,Urbanek2018,Fassbender2009,Obry2013,Wawro2018}. A recent promising concept involves the formation of periodicity by a regular change of magnetization orientation in a homogeneous thin film~\cite{Yu2021}. This is exemplified by periodic stripe domains, which can be considered as one-dimensional MCs~\cite{Banerjee2017,Szulc2022}, or skyrmion lattices, which form 2D MCs~\cite{Diaz2020,Chen2021,Takagi2021,Bassotti2022}. These structures possess a crucial feature of reconfigurable magnetization structure that is sensitive to an external bias magnetic field, allowing the magnonic band structure to be programmed and tuned after device fabrication to suit actual requirements~\cite{Yu2021}. Nevertheless, there is still a lack of understanding of structures that combine both patterning and magnetization texture, especially in the context of SW dynamics.

Recent explorations in magnon-magnon coupling have revealed diverse mechanisms across various systems. In synthetic antiferromagnets and nanomagnonic devices, studies such as \textcite{chen2018strong, sud2020tunable, shiota2020tunable} have demonstrated strong interlayer coupling, influenced by both interlayer exchange and dynamical dipolar interactions, with tunability via external parameters like magnetic field orientation. This coupling manifests as mode splitting and large anticrossing gaps, pivotal for advanced device applications~\cite{li2020hybrid}. Additionally, the work in \textcite{dai2020strong} and \textcite{shiota2020tunable} specifically addresses the coupling between acoustic and optic magnon modes within synthetic antiferromagnets, highlighting the role of bias field tuning and symmetry breaking in achieving strong coupling regimes.
In single-system contexts, such as magnetic skyrmions, different internal mode couplings (e.g., gyrotropic and azimuthal modes) primarily occur through direct exchange interactions, as discussed in \textcite{li2022interaction}. However, the coupling between the spin waves of the two domains, which provides both magnetostatic and exchange interactions mediated by the domain wall, remains unexplored.

In our previous studies, we have shown that multilayer ADLs with perpendicular magnetic anisotropy (PMA) have the potential to be promising patterned systems with periodic magnetization texture to control SW propagation~\cite{Pal2014,Pan2020c,Mantion2022}. Micromagnetic simulations were used to interpret time-resolved magneto-optical Kerr effect microscopy measurements of SW spectra in ADLs based on [Co/Pd]$_8$ multilayers~\cite{Pal2014}. The low-frequency mode observed in the spectra has been attributed to a rim formed during the focused ion beam patterning process at the antidots' edges (see Fig.~\ref{fig1}). The rims exhibit modified magnetic properties due to the area penetrated by Ga$^+$ ions being larger than antidots. Consequently, at remanence, the magnetization assumes an in-plane alignment, generating an MC composed of two regions with different magnetization orientations~\cite{Pan2020c}. This arrangement gives rise to complex SW spectra consisting of bulk modes confined to the ferromagnetic matrix and various modes localized in the rims, with the potential for their hybrid excitations. 
However, the hybridization between the SWs confined to the different regions, especially of different magnetization orientations, of the MC has not been studied so far.

In this paper, we focus on a SW spectrum in antidot lattice with modified rims (ADL-MR) based on [Co/Pd]$_8$ multilayers~\cite{Pan2020c,moalic2022spin} as a function of the strength of an external magnetic field, which is perpendicular to the multilayer plane. We show that magnetic field variation affects bulk- and rim-localized modes in different ways, thereby creating good conditions for bulk and edge mode hybridization. We explain the nature of the bulk-rim interactions resulting in magnon-magnon coupling and formulate conditions for its existence. This discovery unlocks possibilities for hybridizing different modes in ADL, exploring new collective SW phenomena, and advancing practical applications of magnonics.

The structure of the paper is as follows: The next section introduces the system and micromagnetic model utilized in the simulations. Following that, in the Results section, we present the hysteresis of the system, the SW modes in dependence on the magnetic field, the rim-bulk static coupling at low frequencies and the magnon-magnon coupling between second-order radial rim modes and the bulk modes. Finally, we summarize the findings.

\section{Structure and methodology}
\label{Sec:Methods}
\begin{figure}[ht]
    \includegraphics[width=.5\textwidth]{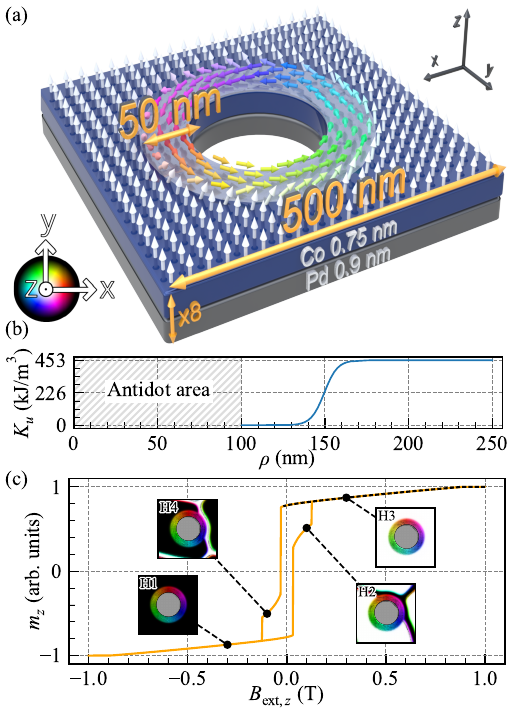}
    \caption{
    (a) Schematic illustration of the investigated structure showing the Co/Pd unit-cell of the ADL-MR. Note that the figure is not to scale. The light gray area around the hole represents the rim with the reduced PMA. The arrows roughly indicate the orientation of the magnetization for $\bextz=0$~T.
    (b) Variation of the PMA constant along the radial direction, starting at the center of the antidot.
    (c) Hysteresis loop along the $z$-axis. The dashed black line marks the regime where the demagnetization process is reversible, and within which the SW spectra are calculated. In the insets, the static magnetization configuration at selected field values is illustrated. The hue represents the in-plane orientation of the magnetization, while the brightness indicates the out-of-plane value, with black being fully down and white fully up. The non-magnetic areas are hatched in gray.
    }
    \label{fig1}
\end{figure}

We are investigating a multilayered [Co/Pd]$_8$ sample containing a square lattice of antidots with a 200 nm diameter, oriented on the $xy$-plane with a lattice constant of 500 nm. 
The magnetic structure is defined with an effective medium approach, where the 8 repetitions of the Co/Pd bilayer stack, consisting of Co (0.75 nm) and Pd (0.9 nm) layers are simulated as a single 13.2 nm-thick layer of Co with effective material parameters~\cite{lemesh2018twisted}. We discretized the system with 256 cuboids along the $x$- and $y$-axes in the square-lattice unit cell shown in Fig.~\ref{fig1}(a), each cuboid being $\approx 1.94 \times 1.94 \times 13.2$~nm$^3$ for a total of $\approx 500 \times 500 \times 13.2$~nm$^3$. For efficiency purposes, we only modeled a single lattice unit cell but we applied periodic boundary conditions with 32 repetitions along the $x$- and $y$-directions to recreate the square lattice.

We assume that each antidot is surrounded by a 50 nm-wide rim, wherein the PMA strength reduces to zero at the antidot edge. 
A hyperbolic tangent function is utilized to model the transition of the PMA value from the bulk value to 0:
$$
K_\text{u}(\rho) = \left(\frac{1}{2} \tanh\left(\frac{\rho - \rho_{\mathrm{edge}}}{8}\right) + \frac{1}{2}\right)  K_{u,{\text{bulk}}} 
$$
for $\rho >$ 100~nm,
where $\rho$ is a radial coordinate relative to the center of antidot, $\rho_{\mathrm{edge}}=150$~nm is the rim edge position, and $K_{u,{\text{bulk}}}$ is the PMA constant of the Co/Pd multilayer, as plotted in Fig.~\ref{fig1}(b).
This function serves as an approximate representation of the anisotropy reduction profile that is likely to be encountered in experimental samples.

For micromagnetic simulations, we use own our version of Mumax3~\cite{mumax_2014,Leliaert2018}, called Amumax~\cite{amumax2023}, which solves the Landau-Lifshitz-Gilbert equation:
\begin{equation}
 \frac{\text{d}\mathbf{m}}{\mathrm{d}t}= 
 \frac{\gamma \mu_0}{1+\alpha^{2}} \left(\mathbf{m} \times \mathbf{H}_{\mathrm{eff}} + 
 \alpha  \mathbf{m} \times 
 (\mathbf{m} \times \mathbf{H}_{\mathrm{eff}}) \right),
\end{equation}
where $\textbf{m} = \textbf{M} / M_{\mathrm{S}}$ is the normalized magnetization, $M_{\mathrm{S}}$ is the magnetization saturation, $\textbf{\text{H}}_{\mathrm{eff}}$ is the effective magnetic field acting on the magnetization, $\gamma=187$ rad/(s$\cdot$T) is the gyromagnetic ratio, $\alpha$ is the damping constant.
The following components were considered in the effective magnetic field $\textbf{H}_{\mathrm{eff}}$: demagnetizing field $\textbf{\text{H}}_{\mathrm{d}}$, exchange field $\textbf{\text{H}}_{\mathrm{exch}}$, uniaxial magnetic anisotropy field  $\textbf{\text{H}}_{\mathrm{anis}}$, and external magnetic field $\textbf{\text{H}}_{\mathrm{ext}}$. Thermal effects were neglected. Thus, the effective field is expressed as:
\begin{equation}
  \textbf{H}_{\mathrm{eff}} =
  \textbf{H}_{\mathrm{d}} + \textbf{H}_{\mathrm{exch}} + \textbf{H}_{\mathrm{ext}} + \textbf{H}_{\mathrm{anis}} +\textbf{h}_{\mathrm{mf}},
\end{equation}
where the last term, $\textbf{h}_{\mathrm{mf}}$ is a microwave magnetic field used for SW excitation. The exchange and anisotropy fields are defined as
\begin{equation}
  \textbf{H}_{\mathrm{exch}} = \frac{2A_{\mathrm{ex}}}{\mu_0 M_{\mathrm{S}}} \Delta \textbf{m},\;
  \textbf{H}_{\mathrm{anis}} =
\frac{2K_{\mathrm{u,bulk}}}{\mu_0 M_{\mathrm{S}}} m_z \hat{\textbf{z}},
\label{Eq:Fields}
\end{equation}
where $A_{\text{ex}}$ is the exchange constant.

We used the following effective magnetic parameters taken from Ref.~[\onlinecite{Pan2020c}] for the Co/Pd multilayer: $K_{\text{u,bulk}} = 4.5\times 10^5$ J/m$^3$, $M_{\mathrm{S}} = 0.81 \times 10^6$ A/m, $A_{\text{ex}} = 1.3 \times 10^{-11}$ J/m. In most simulations, we used a low damping constant $\alpha = 1 \times 10^{-7}$ to get a sharp SW spectrum.

During the simulations, we first relax the magnetization in the system until we reach the ground state. We then excite the SWs with a global microwave magnetic field along the $x$-axis, uniform in space, with a $sinc$ temporal profile and a cut-off frequency of 20 GHz and a peak amplitude of $5 \times 10^4$~T. The excitation field is applied for 1 ns, and we sample the magnetization dynamics at intervals of 16.66 ps over a period of 100 ns.

\begin{figure*}[!t]
    \includegraphics[width=0.99\textwidth]{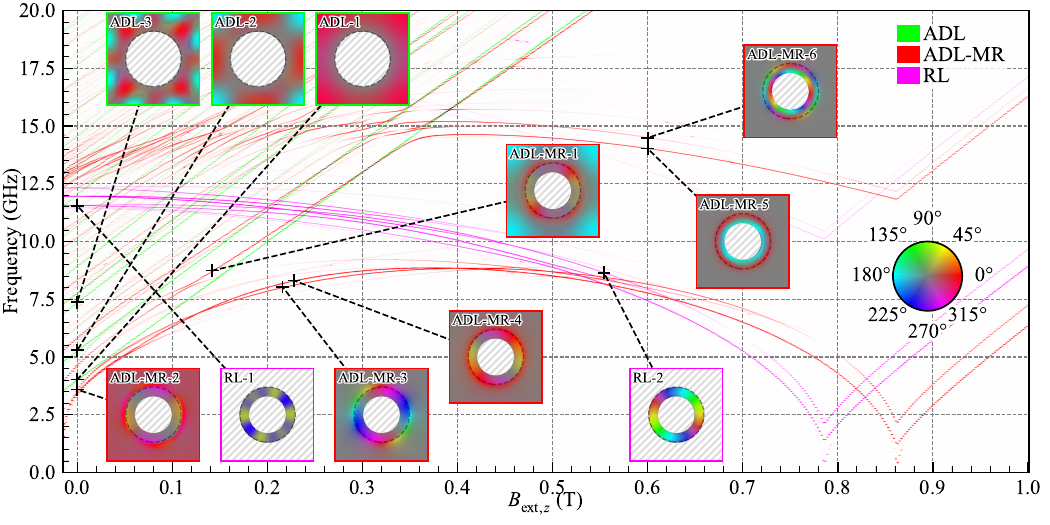}
    \caption{
    Evolution of the SW resonance spectra for the RL, ADL, and ADL-MR in dependence on the external out-of-plane magnetic field. 
    The line intensity correlates with the SW amplitude.
    In the insets, the hue represents the phase of the in-plane dynamic magnetization, while the saturation indicates its spatial amplitude.
    The circular dashed line represents the outer edge of the modified rim. 
    The border color of each inset specifies the corresponding geometry. 
    Non-magnetic areas are hatched in gray.
    A colorblind-accessible version of this figure is available in the supplementary materials.
    }
    \label{fig2}
\end{figure*}

To acquire the SW spectrum, we took the space- and time-resolved in-plane magnetization and applied a Hanning window along the time axis. We then computed the real discrete Fourier transform using the fast Fourier transform (FFT) algorithm along the time axis for each cuboid composing the system. After this process, for every discrete frequency, we pinpointed the cell exhibiting the maximum amplitude. This step was repeated iteratively for each frequency to progressively construct the spectrum. 
By selecting the highest amplitude rather than the average, the strongly-localized mode is emphasized over the modes that are spread over a wide area, such as the bulk modes. This procedure is applied to each simulation, where the external magnetic field value varies. The SW spectral response (Figs.~\ref{fig2},~\ref{fig3}, \ref{fig4}, and \ref{fig5}) in dependence on the external magnetic field directed along the $z$-axis were calculated for values from 1 T to $-$0.014 T in decrements of 2 mT. To prevent numerical artifacts arising from a super-symmetry of the spins, we angle the external field by 0.0001 degrees from the $z$-axis.
To generate the mode visualizations, we separately took each cuboid making up the system and we calculated the FFT of the in-plane magnetization over time. Then, for a selected frequency, we map the modulus of the complex number to a saturation value between 0 and 1 and the argument of the complex number to a hue where it is red if the argument is 0. This process is repeated for each cuboid in the system.

\section{Results}

\subsection{Hysteresis}

First, we study the change of the static magnetization configuration in ADL-MR with the magnetic field applied out of the plane. The hysteresis loop along this direction, changing the magnetic field intensity from $\bextz= \mu_0 H_{\mathrm{ext},z}= -1$~T to 1~T and back is shown in Fig.~\ref{fig1}(c). At remanence, the bulk magnetization keeps an out-of-plane saturation, while in the rim area, it stabilizes into a vortex-like state. Upon switching the external magnetic field, domain walls emerge within the bulk starting at $\bextz= \pm 0.014$ (see, states \textit{H2} and \textit{H4}). Consequently, at field $\bextz=\pm 0.124$ T, the domain evolves to full saturation in the opposite direction (\textit{H1} and \textit{H3}). The rim magnetization reaches full saturation when the magnetic field intensity exceeds $\bextz=0.86$ T or falls below $\bextz=-0.86$ T, resulting in an out-of-plane saturation. In the following section, we will examine the SW dynamics in the field range marked by the dashed line in Fig.~\ref{fig1}(c) (i.e., between 1 and $-0.014$ T), which signifies the full saturation of the bulk. 
The clockwise (CW) and counter-clockwise (CCW) vortex-like states in the rims are degenerated, therefore this chirality do not affect the frequencies of the azimuthal modes within the rims. For this reason, when not saturated out of the plane, the magnetization in the rims was chosen to be in the CCW vortex-like state in the rest of this paper.

\subsection{SW modes in dependence on the magnetic field}

Before examining the field dependence of SWs in ADL-MR, we analyze the spectra of two complementary subsystems: a square lattice of rings (RL) with a width of 50 nm but without PMA, plotted in purple in Fig.~\ref{fig2}; and a simplified ADL utilizing PMA but without rims, with an antidot diameter of 300 nm plotted in green in Fig.~\ref{fig2} and marked as ADL. The other geometric and material parameters for subsystems remain the same as in ADL-MR. The spatial amplitude distribution of the SW modes is presented in insets adjacent to the spectra. The color of each inset's boundary and label indicates its affiliation with a specific system. Colorblind-accessible visualizations of the ADL, RL and ADL-MR spectra are provided in the supplementary materials in Fig.~S1, S3, and S2 respectively and a detailed visualization of all the modes from Fig.~\ref{fig2} are displayed in Fig.~S7.
At remanence in the ADL system, we found the fundamental mode \textit{ADL-1} (the mode numbering refers to the order of modes appearing in the discussion) at $f=4.02$~GHz and a multitude of higher-order bulk modes, such as \textit{ADL-2} and \textit{ADL-3}, at $f=5.29$ and 7.37 GHz, respectively. The system is characterized by a standard linear relationship between the SW frequency and the external field~\cite{Pal2014}. Therefore, as the field increases, no interaction between the modes is observed. 

In the RL system with a CCW vortex-like magnetization texture, all the modes have an azimuthal component. At $\bextz=0$ T, the three curves of lowest frequency are at $f=11.55$ (\textit{RL-1}), 11.94 and 12.32 GHz, and correspond to the 3rd, 1st and 5th order azimuthal modes, respectively. These bands result from the degeneration of CW and CCW azimuthal modes of the same order. 
Increasing the magnetic field results in the splitting of this degeneration~\cite{Dugaev2005} and a monotonous decrease in the frequency of the modes, until the magnetization saturates out of plane for 0.78 T. The second-order radial modes start at around 17.5 GHz at remanence and decrease frequency to 10 GHz at the magnetization reorientation field. All SWs increase in frequency as the magnetic field continues to increase above 0.78 T.

The SW spectrum as a function of external magnetic field for ADL-MR is plotted in red in Fig.~\ref{fig2}. 
By comparison with the spectra of ADL and RL, we can distinguish two main groups of branches that are related to: (i) the bulk modes that are similar to those in the ADL system, and exhibit a linear frequency increase trend with an increasing magnetic field (compare the branches \textit{ADL-MR-1} and \textit{ADL-1}); (ii) rim modes, like \textit{ADL-MR-2}, \textit{ADL-MR-3}, and \textit{ADL-MR-4}, which exhibit a non-monotonic frequency response to changes in the applied magnetic field. Rim modes are similar to those in the RL system at large fields, $\bextz \gtrsim 0.4$~T (e.g., \textit{ADL-MR-2} is analogous to \textit{RL-2}). 
Beyond $\bextz=0.4$ T, these branches decrease in frequency due to the SW mode softening until they reach a phase transition at $\bextz=0.862$ T. Subsequently, the frequency of these branches increases as the magnetic field strength increases further.

Although the behavior of the two complementary systems, ADL and RL, merge in the ADL-MR system's spectra, some effects remain distinct to ADL-MR. Specifically, we can identify two frequency ranges where collective bulk-rim effects occur. The first range (i) spans low frequencies from 3 to 12.5 GHz, characterized by first-order radial modes in the rims, i.e., modes whose phase does not change sign along the radial direction in the rim (e.g., \textit{ADL-MR-2}, \textit{ADL-MR-3}, and \textit{ADL-MR-4}).

The second range (ii) 12.5-17.5 GHz is where the second-order radial modes are in the rim (e.g., \textit{ADL-MR-5} and \textit{ADL-MR-6}). In (i), the bulk modes dependence on $\bextz$ deviates from a straight-line relationship (see \textit{ADL-MR-1}), and the branches of the rim modes (e.g., \textit{ADL-MR-2} and \textit{ADL-MR-3}) gain dependencies similar to the bulk modes at low field values, leading to a decrease in frequency as the magnetic field decreases starting at $\bextz \lesssim 0.4$~T. In turn, regarding the (ii) range, there are multiple points of intersection between the bulk and rim modes resulting in anti-crossings. 
 
The aforementioned effects of ADL-MR indicate a magnetic coupling between bulk and rim, which is the focus of this paper. However, these effects are influenced by multiple factors: gradual change of the magnetization orientation within the rims from an in-plane to an out-of-plane configuration with $\bextz$ change (see the hysteresis loop in Fig.~\ref{fig1}(a)), change in the magnetization gradient between rim and bulk, static stray magnetic field of the bulk area on the rim, and finally dynamical coupling between bulk and rim SW modes. In the next two sections, the effects found in the ADL-MR are discussed individually for the first- and second-order radial rim modes, i.e., at low (3-12.5~GHz) and high (12.5-17.5~GHz) frequency ranges. 

\subsection{Rim-bulk coupling at low frequencies: first-order radial modes}

\begin{figure}[ht]
    \includegraphics[width=.5\textwidth]{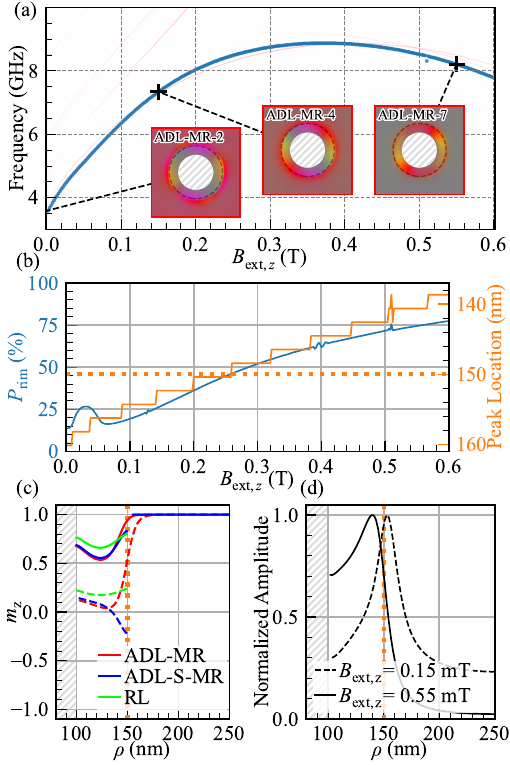}
    \caption{
    (a) Evolution of the SW resonance spectra for the ADL-MR in dependence on the external out-of-plane magnetic field. 
    The line intensity correlates with the SW amplitude.
    The branch highlighted in blue corresponds to the first-order CW azimuthal mode considered in (b) and (d).
    The static configuration in (c) and the modes in (d) are marked by black crosses.
    (b) Rim/bulk mode intensity ratio (in blue) and peak maximum location (in orange) in dependence on the external out-of-plane magnetic field.
    (c) Profile of the out-of-plane static magnetization for $\bextz=0.150$~T (dashed lines) and $\bextz=0.550$~T (solid lines) for 3 systems (ADL-MR, ADL-S-MR and RL).
    (d) Amplitude profile of the marked modes from (a).
    The orange dotted line indicates the boundary between the rim and the bulk.
    }
    \label{fig3}
\end{figure}

Our results have shown that the frequencies of the first-order radial modes in the rim are significantly influenced by the interaction with the bulk part of the ADL-MR. The effects of this interaction are also visible in the azimuthal mode profiles shown in Fig.~\ref{fig2}, \textit{ADL-MR-2, 3, 4}. They show the transition of the mode localization, where at low magnetic field amplitudes, the mode is predominantly located in the bulk (\textit{ADL-MR-2}) and then gradually shifts its position towards the edge region (\textit{ADL-MR-4}). To explain this bulk-rim coupling, our subsequent analysis will focus on the first-order CW azimuthal mode (i.e., the mode where the $m_z$ component undergoes a phase change of $2\pi$ along the path encircling the rim, as illustrated in Figs. S7-S10 in the supplementary material). This mode is highlighted in blue in Fig.~\ref{fig3}(a), where profiles at three selected values of $\bextz$, namely \textit{ADL-MR-2}, \textit{ADL-MR-4}, and \textit{ADL-MR-7}, are also depicted.A detailed visualization of all the modes from Fig.~\ref{fig3}(a) are displayed in Fig.~S8 in the supplementary materials.

We evaluate the spatial distribution of this mode's amplitude, particularly its concentration in bulk or rim. To quantify the extent of concentration, we define the parameter $\Prim = \Arim / \Aunit$, where $\Arim$ and $\Aunit$ denote the integrated amplitude profiles within the rim and the unit cell defined as $ \Arim = \int_{100}^{150} \int_{0}^{2\pi} \mathrm{FFT(\mathbf{m}(\rho))^2} \, d\rho d\phi$ and $  \Aunit = \int_{0}^{500}\int_{0}^{500} \mathrm{FFT(\mathbf{m}(}x,y))^2 \, dxdy $ with the integration over the surface of the rim and the surface of the unit cell excluding the antidot, respectively.
$\Prim$ as a function of $\bextz$ is visualized in blue in Fig.~\ref{fig3}(b).
A $\Prim$ value of 0 and 100\% indicates that the mode is entirely concentrated within the bulk and rim, respectively. At remanence, the mode concentrates predominantly in the bulk as $\Prim=14\%$, the intensity is equally distributed between the rim and the bulk at 0.282 T, and, at higher fields, this mode is predominantly concentrated in the rim.

We also quantify the shift of the amplitude maximum of this mode as a function of $\bextz$, as shown in orange in Fig.~\ref{fig3}(b). Starting at 160 nm at remanence, i.e., in the bulk of the ADL-MR, the position of the peak moves towards the rim as the magnetic field increases. It reaches the rim edge (150 nm) at 0.2 T and moves into the rim region at higher fields. Confirmation of this behavior is shown in Fig.~\ref{fig3}(d), which shows the normalized amplitude profiles of the mode, marked by black crosses in Fig.~\ref{fig3}(a), at $\bextz$ values of 0.15 and 0.55 T. As $\bextz$ is increased, not only does the amplitude maximum of the mode move to the rim, but the profile undergoes a transformation, becoming less localized and exhibiting a flattened distribution within the rim region.

\begin{figure*}[!t]
    \includegraphics[width=0.99\textwidth]{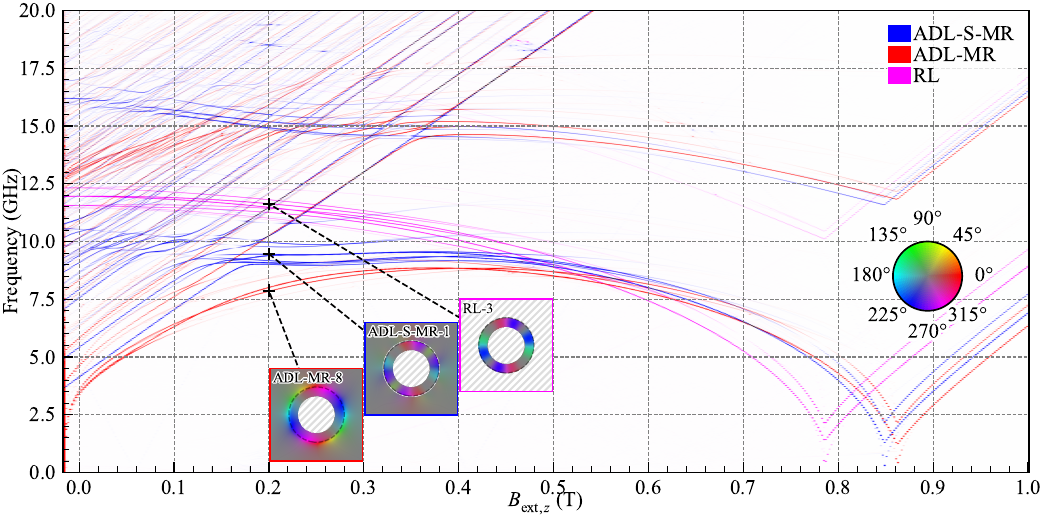}
    \caption{
    Evolution of the SW resonance spectra for the ADL-MR, ADL-S-MR, and RL in dependence on the external out-of-plane magnetic field. 
    The line intensity correlates with the SW amplitude.
    In the insets, the hue represents the phase of the in-plane dynamic magnetization, while the saturation indicates its spatial amplitude.
    The circular dashed line represents the outer edge of the modified rim. 
    The border color of each inset specifies the corresponding geometry. 
    Non-magnetic areas are hatched in gray.
    A colorblind-accessible version of this figure is available in the supplementary materials.
    }
    \label{fig4}
\end{figure*}

As $\bextz$ is increased, not only both the mode profile and the mode intensity's maximum location shift towards the rim area but there is also a concomitant change in the static magnetization in the rim. The radial cross-section of the $m_{z}$ component of the static magnetization, starting from the antidot edge ($\rho=100$~nm), is presented for the ADL-MR and the RL configurations in Fig.~\ref{fig3}(c). The dependencies are depicted for two distinct external field strengths, $\bextz=0.15$ T (dashed lines) and $\bextz=0.55$ T (solid lines). Notably, the RL configuration exhibits a larger out-of-plane magnetization component compared to the other structures, attributable to the absence of a static stray magnetic field from the bulk. Moreover, the large magnetization gradient between the bulk and rim shifts towards the rim area. The gradient strength reduces with the increase of the external magnetic field due to decreasing difference between the out-of-plane magnetization component in the rim and bulk. The position of the mode intensity maximum agrees with the position of the largest magnetization gradient.

The consequence of changing the magnetization texture at small $\bextz$ is a change in the location and concentration of the azimuthal modes, discussed above (see also Fig.~\ref{fig3}(b-d)), resulting also in their significant frequency reduction compared to the RL reference system. The mode softening is an effect characteristic of systems with large magnetization gradients like domain walls, vortices, and skyrmions~\cite{varentcova2020toward}. In a region with a large magnetization gradient, the effective field and energy are reduced (due to compensation of the exchange and anisotropy contributions) so that the magnetic moment requires less energy to precess. It results in a significant reduction of the frequency.

To further pinpoint the origins of the observed rim-bulk interactions shown in Figs.~\ref{fig2}-\ref{fig3}, we conducted simulations of the ADL-MR with a 5.82 nm-wide non-magnetic spacer between the rim and the bulk (ADL-S-MR). The SW spectrum in dependence on $\bextz$ is depicted in blue in Fig.~\ref{fig4}. Colorblind-accessible visualizations of the ADL-S-MR, RL, and ADL-MR spectra are provided in  Figs.~S4, S3, and S2, respectively, and a detailed visualization of all the modes from Fig.~\ref{fig4} are displayed in Fig.~S9 in the supplementary materials. This approach effectively nullifies the exchange interaction between the rim and the bulk, thereby isolating the dipolar interaction as the only carrier of the rim-bulk interactions. Consequently, the introduction of this spacer eliminates the large magnetization gradient that existed between the bulk and the rim in ADL-MR (see Fig.~\ref{fig3}(c)). However, the magnetic configuration inside the rim is different from that of the RL. This is due to the stray field from the bulk.

    \begin{figure*}[!t]
        \includegraphics[width=0.99\textwidth]{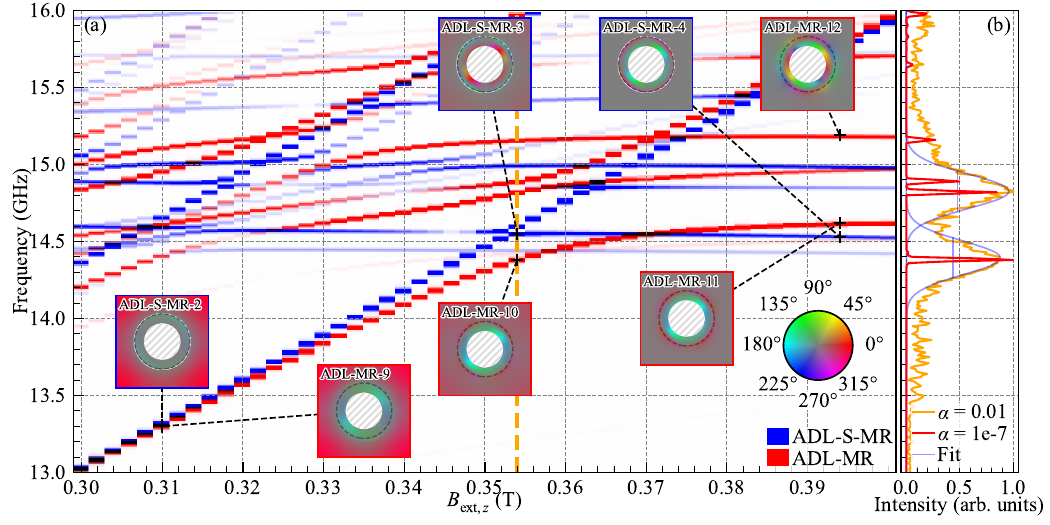}
        \caption{
        (a) A zoom-in part of Fig.~\ref{fig4}, which shows the evolution of the SW resonance spectra for the ADL-MR and ADL-S-MR in dependence on the external out-of-plane magnetic field at a higher frequency range. 
        The orange dashed vertical line indicated the value for which (b) was plotted.
        The line intensity correlates with the SW amplitude.
        In the insets, the hue represents the phase of the in-plane dynamic magnetization, while the saturation indicates its spatial amplitude.
        The circular dashed line represents the outer edge of the modified rim. 
        The border color of each inset specifies the corresponding geometry. 
        Non-magnetic areas are hatched in gray.
        (b) Ferromagnetic resonance spectra of ADL-MR for \bextz~=~0.356~T with low damping in red and realistic damping ($\alpha = 0.01$) in orange. The blue line is a fit of the spectra shown in orange with two Gaussian curves.
        A colorblind-accessible version of this figure is available in the supplementary materials.
        }
        \label{fig5}
    \end{figure*}

Within the field range from $\bextz=0$ to 0.5~T, the ADL-S-MR system exhibits higher frequencies relative to those in the ADL-MR system, and their field dependence resembles that of the RL system, although with lower frequencies. Interestingly, the azimuthal modes of the rim in the ADL-S-MR system couples strongly with the bulk modes, which is indicated by anti-crossings (see range 8.5-10.5 GHz in Fig.~\ref{fig4}). This indicates that the exchange interaction between the bulk and the rim in ADL-MR has a significant impact on the azimuthal modes of the rim. 

We select the first-order azimuthal CW modes at 0.2~T for the ADL-MR, ADL-S-MR, and RL systems, it is at 7.87, 9.46, and 11.63~GHz, respectively, to visualize their amplitude distribution, which is shown in Fig.~\ref{fig4} in the insets with labels \textit{ADL-MR-8}, \textit{ADL-S-MR-1} and \textit{RL-3}, respectively. These modes and their CCW counterparts are the most intense and robust branches as they have a strong intensity even for higher fields and after saturation. Comparing \textit{ADL-MR-8} with \textit{ADL-S-MR-1}, we see higher SW amplitude in the bulk in ADL-MR system, which indicates that the exchange coupling between the rim and the bulk may increase the dynamic coupling between the rim and bulk magnons.

The observations discussed above show that the presence of a static magnetization gradient, its evolution, and the reorientation of the magnetization in the rim with a change in the magnetic field intensity are the most important factors influencing the SW spectra, in particular, mode softening in ADL-MR at low frequencies and low magnetic fields. On top of this, there is a dynamical magnon-magnon coupling, which is clearly evidenced by the amplitude profiles of the predominantly bulk (\textit{ADL-MR-1} and \textit{ADL-MR-2}) and rim (\textit{ADL-MR-4}) modes, which have properties of both the fundamental bulk mode and the first-order azimuthal CW mode (see Fig. ~\ref{fig2} and S7 in the supplementary materials). However, the overlap of the static magnetization deformation and the dynamical coupling change with the external magnetic field makes the estimation of the dynamical coupling strength difficult. 

\subsection{Magnon-magnon coupling between second-order radial rim modes and the bulk modes}

Let us now move on to the analysis of the spectrum in the higher frequency range, where the interaction between the second-order radial rim modes and the bulk modes exists. The detailed spectra are shown in Fig.~\ref{fig5}, which is a zoom-in part of Fig.~\ref{fig4}. Colorblind-accessible visualizations of the zoomed in ADL-MR and ADL-S-MR spectra are provided in Figs.~S5 and S6 respectively and a detailed visualization of all the modes from Fig.~\ref{fig5}(a) are displayed in Fig.~S10 in the supplementary materials. Here, the horizontal branches are related to the azimuthal SW modes concentrated in the rim, which have a $\pi$ phase change along the radial direction (see the insets and figures in the supplementary material), while the lines with a nonzero slope originate from the bulk modes of the ADL. Comparing the spectra of ADL-MR (in red) with the ADL-S-MR (in blue), it is evident that the ADL-MR exhibits a number of hybridizations between selected bulk modes and second-order radial rim modes, as evidenced by large gaps between the branches, and a strong deviation from the linear dependencies $f(\bextz)$. Conversely, the ADL-S-MR system demonstrates crossings with only tiny gaps, especially between fundamental bulk mode (blue line along the diagonal of the plot in Fig.~\ref{fig5}) and second-order radial rim modes, as indicated by vertical orange line at $\bextz = 0.354$ T. Considering that the only difference between ADL-MR and ADL-S-MR systems is the break of the exchange interaction between the rim and the bulk of the ADL in the latter, we can conclude that the exchange interaction is responsible for the strong magnon-magnon coupling observed in the former system. Nevertheless, it is still the hybrid coupling in ADL-MR, involving dynamical magnon-magnon coupling, dominating in this range of fields and frequencies but modified by the magnetization texture with a change of $\bextz$. 

We aim to understand why some rim and bulk modes show large hybridization and some do not in ADL-MR. To find the matching condition, we compare the \textit{ADL-MR-9/10/11} modes shown in Fig.~\ref{fig5}(a), which correspond to the bulk mode, bulk+rim mode, and rim mode, respectively. From there, we can see the smooth transition of the amplitude of the mode as we increase the field. From \textit{ADL-MR-9} to \textit{ADL-MR-10}, we see the bulk part of the mode which is first outside of the rim for $\bextz=0.31$~T to partly inside the rim for $\bextz=0.354$~T while the rim amplitude increases. The azimuthal homogeneous nature of the outer ring of the rim mode can still be found at $\bextz=0.394$~T for \textit{ADL-MR-11} while the amplitude in the bulk is close to null. In comparison, \textit{ADL-MR-12} has a second-order quantization in the azimuthal direction and we observe the crossing of its branch and a fundamental bulk mode at $\bextz=0.37$~T for $f=15.16$~GHz. A deeper analysis of the mode profiles indicates the capability of coupling between rim and bulk modes if the number of azimuthal order of these modes is the same after modulo 4 operation. Such a nature of coupling is strictly connected with the fact that the symmetry of bulk modes is governed by the fourfold symmetry of the square lattice.

Additionally, for ADL-S-MR, we can find a much reduced hybridization length of 50 MHz for $\bextz= 0.352$~T as opposed to 380~MHz for ADL-MR. Similarly, as for ADL-MR, we can look at the evolution of the mode profile as we increase the field. The modes \textit{ADL-MR-9/10/11} are taken to be analogous to \textit{ADL-S-MR-2/3/4} for the same branches and the same fields. Removing the exchange interaction between the rim and the bulk had an effect on the bulk mode expanding into the rim area, which is not present in the middle of the hybridization on \textit{ADL-S-MR-3} like it is for \textit{ADL-MR-10}. We can then confirm that the exchange interaction is responsible for this particular hybridization.

In Fig.~\ref{fig5}(b) showing the ferromagnetic resonance spectrum at 0.354~T, we demonstrate that the hybridization under consideration in the ADL-MR system is still visible when using a realistic damping value of $\alpha=0.01$~\cite{Pan2020c}, which is close to the experimental value for Co/Pd multilayers. We fitted the two resonant modes of frequencies $f=14.39$ and 14.83~GHz with the Gaussian curves. 
Based on this, we estimate the cooperativity of the bulk magnon-rim magnon coupling. The strength of the coupling, $g$, is defined as half of the minimal peak-to-peak frequency spacing in the anti-crossing, it is $g=215$~MHz at 0.354~T. The half width at half maximum for both peaks is $\kappa_\text{low-f}=128$ MHz and $\kappa_\text{high-f}=131$ MHz. Using these values, we calculate the cooperativity
$$
C=\frac{g^2}{\kappa_\text{low-f} \times \kappa_\text{high-f}}= 2.757.
$$
Even though the system was not optimized for it, this value of the cooperativity indicates a strong magnon-magnon coupling in the ADL-MR between the second-order radial, first-order azimuthal rim mode and the fundamental bulk ADL mode. 

We can compare this cooperativity with the values given for various systems presented in the literature. For instance, Chen et al.~\cite{chen2018strong} report $C=0.38$ and $C=21$ in ferromagnetic metallic nanowires magnetized in parallel or antiparallel order, respectively, which are forming an array deposited on YIG film. 
%MacNeill et al. \cite{macneill2019gigahertz} find C=6.4$ in antiferromagnetic CrCl3 layered crystal with ferromagnetic order in each layer.
In the single planar ferromagnetic nanoelement with optimized ends, Dai et al. \cite{dai2020strong} achieved $C=60.1$ for the hybridization between bulk and the edge SW modes. Adhikari et al. \cite{adhikari2020large} reports the coupling between magnons in Ni$_{80}$Fe$_{20}$ nanocross with cooperativity $C=0.28$, which can be enhanced to 2.5 by making a hole in the nanocross~\cite{adhikari2021observation}. Synthetic antiferromagnets are also considered for exploitation of the magnon-magnon coupling. Here, the coupling strength depends directly on the RKKY interaction strength between ferromagnetic layers (controlled by the thickness of the non-magnetic interlayer) but also external magnetic field which changes the relative orientation of the magnetizations in the layers. Depending on materials used, the reported values are $C=25.0$~\cite{dai2021strong}, $C=5.26$~\cite{ma2023strong} or $C=8.4$~\cite{hayashi2023observation}. Just recently, Dion et al. \cite{dion2023ultrastrong} achieved probably the largest magnon-magnon cooperativity so far $C=126.4$ in finite-size magnetostatically-coupled ferromagnetic bilayers, thus combining inter-element coupling and the final size to enhance the coupling between magnons in both layers. 
In the context presented, the coupling between rim and bulk modes in ADL-MR system based on PMA material, as demonstrated above, explores a new type of rather strong dynamic coupling between planar regions of non-collinear magnetization, which is mediated mainly by exchange interactions and turns on a higher-order azimuthal mode. The influence of the lattice type, as indicated by the hybridization selection rules, suggests a possibility for further optimization of the coupling strength, not only by changing the material but also by changing the geometry of the ADL. 

\section{Conclusions}

Using micromagnetic simulations, we study the SW dynamics in a thin film PMA magnonic crystal consisting of antidots with rims around the antidot edges with in-plane magnetization (ADL-MR) as a function of the out-of-plane magnetic field strength. Through comparative studies with its complementary subsystems, i.e. pure ADL and the rim lattice, we have shown that the interactions between the subsystems significantly modify the spectrum of the collective SWs and its dependence on the magnetic field. The study reveals SW hybridizations between different waves concentrated in the rim and the bulk,  which depend on the mode type and the out-of-plane magnetic field strength and are additionally modified by the associated changes in the magnetization state at the interface between the subsystems. 

We show that in our system a strong magnon-magnon coupling between the fundamental bulk mode and the second-order radial azimuthal mode is present. It is characterized by a cooperativity of \( C=2.757 \) which is competitive compared to the other magnon-magnon- couplings reported so far in the literature. We show that this coupling is mainly determined by the exchange interactions between the bulk and the rim, which are mediated by the domain wall. In addition, it is influenced by the lattice type, which in combination with the lattice and material parameters opens a wide field for further optimization and exploitation of the collective hybrid dynamics in such systems. Thus, the demonstrated properties are pivotal for the development of magnonic applications and devices, including hybrid and quantum magnonics.

\section*{Acknowledgments}
The research has received funding from the National Science Centre of Poland, Grant No. UMO–2020/37/B/ST3/03936 and 2023/49/N/ST3/03538. The simulations were partially performed at the Poznan Supercomputing and Networking Center (Grant No. 398).

\bibliography{ref}

\begin{thebibliography}{50}%
\makeatletter
\providecommand \@ifxundefined [1]{%
 \@ifx{#1\undefined}
}%
\providecommand \@ifnum [1]{%
 \ifnum #1\expandafter \@firstoftwo
 \else \expandafter \@secondoftwo
 \fi
}%
\providecommand \@ifx [1]{%
 \ifx #1\expandafter \@firstoftwo
 \else \expandafter \@secondoftwo
 \fi
}%
\providecommand \natexlab [1]{#1}%
\providecommand \enquote  [1]{``#1''}%
\providecommand \bibnamefont  [1]{#1}%
\providecommand \bibfnamefont [1]{#1}%
\providecommand \citenamefont [1]{#1}%
\providecommand \href@noop [0]{\@secondoftwo}%
\providecommand \href [0]{\begingroup \@sanitize@url \@href}%
\providecommand \@href[1]{\@@startlink{#1}\@@href}%
\providecommand \@@href[1]{\endgroup#1\@@endlink}%
\providecommand \@sanitize@url [0]{\catcode `\\12\catcode `\$12\catcode `\&12\catcode `\#12\catcode `\^12\catcode `\_12\catcode `\%12\relax}%
\providecommand \@@startlink[1]{}%
\providecommand \@@endlink[0]{}%
\providecommand \url  [0]{\begingroup\@sanitize@url \@url }%
\providecommand \@url [1]{\endgroup\@href {#1}{\urlprefix }}%
\providecommand \urlprefix  [0]{URL }%
\providecommand \Eprint [0]{\href }%
\providecommand \doibase [0]{https://doi.org/}%
\providecommand \selectlanguage [0]{\@gobble}%
\providecommand \bibinfo  [0]{\@secondoftwo}%
\providecommand \bibfield  [0]{\@secondoftwo}%
\providecommand \translation [1]{[#1]}%
\providecommand \BibitemOpen [0]{}%
\providecommand \bibitemStop [0]{}%
\providecommand \bibitemNoStop [0]{.\EOS\space}%
\providecommand \EOS [0]{\spacefactor3000\relax}%
\providecommand \BibitemShut  [1]{\csname bibitem#1\endcsname}%
\let\auto@bib@innerbib\@empty
%</preamble>
\bibitem [{\citenamefont {Gulyaev}\ and\ \citenamefont {Nikitov}(2001)}]{Gulyaev2001}%
  \BibitemOpen
  \bibfield  {author} {\bibinfo {author} {\bibfnamefont {Y.~V.}\ \bibnamefont {Gulyaev}}\ and\ \bibinfo {author} {\bibfnamefont {S.~A.}\ \bibnamefont {Nikitov}},\ }\href@noop {} {\bibfield  {journal} {\bibinfo  {journal} {Dokl. Phys.}\ }\textbf {\bibinfo {volume} {46}},\ \bibinfo {pages} {687} (\bibinfo {year} {2001})}\BibitemShut {NoStop}%
\bibitem [{\citenamefont {Puszkarski}\ and\ \citenamefont {Krawczyk}(2003)}]{Puszkarski03}%
  \BibitemOpen
  \bibfield  {author} {\bibinfo {author} {\bibfnamefont {H.}~\bibnamefont {Puszkarski}}\ and\ \bibinfo {author} {\bibfnamefont {M.}~\bibnamefont {Krawczyk}},\ }\href@noop {} {\bibfield  {journal} {\bibinfo  {journal} {Solid State Phenom.}\ }\textbf {\bibinfo {volume} {94}},\ \bibinfo {pages} {125} (\bibinfo {year} {2003})}\BibitemShut {NoStop}%
\bibitem [{\citenamefont {Lenk}\ \emph {et~al.}(2011)\citenamefont {Lenk}, \citenamefont {Ulrichs}, \citenamefont {Garbs},\ and\ \citenamefont {M{\"{u}}nzenberg}}]{Lenk2011}%
  \BibitemOpen
  \bibfield  {author} {\bibinfo {author} {\bibfnamefont {B.}~\bibnamefont {Lenk}}, \bibinfo {author} {\bibfnamefont {H.}~\bibnamefont {Ulrichs}}, \bibinfo {author} {\bibfnamefont {F.}~\bibnamefont {Garbs}},\ and\ \bibinfo {author} {\bibfnamefont {M.}~\bibnamefont {M{\"{u}}nzenberg}},\ }\href {https://doi.org/10.1016/j.physrep.2011.06.003} {\bibfield  {journal} {\bibinfo  {journal} {Phys. Rep.}\ }\textbf {\bibinfo {volume} {507}},\ \bibinfo {pages} {107} (\bibinfo {year} {2011})}\BibitemShut {NoStop}%
\bibitem [{\citenamefont {Krawczyk}\ and\ \citenamefont {Grundler}(2014)}]{Krawczyk14}%
  \BibitemOpen
  \bibfield  {author} {\bibinfo {author} {\bibfnamefont {M.}~\bibnamefont {Krawczyk}}\ and\ \bibinfo {author} {\bibfnamefont {D.}~\bibnamefont {Grundler}},\ }\href@noop {} {\bibfield  {journal} {\bibinfo  {journal} {J. Phys. Cond. Matter}\ }\textbf {\bibinfo {volume} {26}},\ \bibinfo {pages} {123202} (\bibinfo {year} {2014})}\BibitemShut {NoStop}%
\bibitem [{\citenamefont {Nikitov}\ \emph {et~al.}(2015)\citenamefont {Nikitov}, \citenamefont {Kalyabin}, \citenamefont {Lisenkov}, \citenamefont {Slavin}, \citenamefont {Barabanenkov}, \citenamefont {Osokin}, \citenamefont {Sadovnikov}, \citenamefont {Beginin}, \citenamefont {Morozova}, \citenamefont {Filimonov}, \citenamefont {Khivintsev}, \citenamefont {Vysotsky}, \citenamefont {Sakharov},\ and\ \citenamefont {Pavlov}}]{Nikitov2015}%
  \BibitemOpen
  \bibfield  {author} {\bibinfo {author} {\bibfnamefont {S.~A.}\ \bibnamefont {Nikitov}}, \bibinfo {author} {\bibfnamefont {D.~V.}\ \bibnamefont {Kalyabin}}, \bibinfo {author} {\bibfnamefont {I.~V.}\ \bibnamefont {Lisenkov}}, \bibinfo {author} {\bibfnamefont {A.}~\bibnamefont {Slavin}}, \bibinfo {author} {\bibfnamefont {Y.~N.}\ \bibnamefont {Barabanenkov}}, \bibinfo {author} {\bibfnamefont {S.~A.}\ \bibnamefont {Osokin}}, \bibinfo {author} {\bibfnamefont {A.~V.}\ \bibnamefont {Sadovnikov}}, \bibinfo {author} {\bibfnamefont {E.~N.}\ \bibnamefont {Beginin}}, \bibinfo {author} {\bibfnamefont {M.~A.}\ \bibnamefont {Morozova}}, \bibinfo {author} {\bibfnamefont {Y.~A.}\ \bibnamefont {Filimonov}}, \bibinfo {author} {\bibfnamefont {Y.~V.}\ \bibnamefont {Khivintsev}}, \bibinfo {author} {\bibfnamefont {S.~L.}\ \bibnamefont {Vysotsky}}, \bibinfo {author} {\bibfnamefont {V.~K.}\ \bibnamefont {Sakharov}},\ and\ \bibinfo {author} {\bibfnamefont {E.~S.}\ \bibnamefont {Pavlov}},\ }\href
  {http://stacks.iop.org/1063-7869/58/i=10/a=1002} {\bibfield  {journal} {\bibinfo  {journal} {Phys. Usp.}\ }\textbf {\bibinfo {volume} {58}},\ \bibinfo {pages} {1002} (\bibinfo {year} {2015})}\BibitemShut {NoStop}%
\bibitem [{\citenamefont {Rych{\l}y}\ \emph {et~al.}(2015)\citenamefont {Rych{\l}y}, \citenamefont {Gruszecki}, \citenamefont {Mruczkiewicz}, \citenamefont {K{\l}os}, \citenamefont {Mamica},\ and\ \citenamefont {Krawczyk}}]{Rychly2015}%
  \BibitemOpen
  \bibfield  {author} {\bibinfo {author} {\bibfnamefont {J.}~\bibnamefont {Rych{\l}y}}, \bibinfo {author} {\bibfnamefont {P.}~\bibnamefont {Gruszecki}}, \bibinfo {author} {\bibfnamefont {M.}~\bibnamefont {Mruczkiewicz}}, \bibinfo {author} {\bibfnamefont {J.~W.}\ \bibnamefont {K{\l}os}}, \bibinfo {author} {\bibfnamefont {S.}~\bibnamefont {Mamica}},\ and\ \bibinfo {author} {\bibfnamefont {M.}~\bibnamefont {Krawczyk}},\ }\href {https://doi.org/10.1063/1.4932348} {\bibfield  {journal} {\bibinfo  {journal} {Low Temp. Phys.}\ }\textbf {\bibinfo {volume} {41}},\ \bibinfo {pages} {745} (\bibinfo {year} {2015})}\BibitemShut {NoStop}%
\bibitem [{\citenamefont {Chumak}\ \emph {et~al.}(2017)\citenamefont {Chumak}, \citenamefont {Serga},\ and\ \citenamefont {Hillebrands}}]{Chumak2017}%
  \BibitemOpen
  \bibfield  {author} {\bibinfo {author} {\bibfnamefont {A.~V.}\ \bibnamefont {Chumak}}, \bibinfo {author} {\bibfnamefont {A.~A.}\ \bibnamefont {Serga}},\ and\ \bibinfo {author} {\bibfnamefont {B.}~\bibnamefont {Hillebrands}},\ }\href {https://doi.org/10.1088/1361-6463/AA6A65} {\bibfield  {journal} {\bibinfo  {journal} {J. Phys. D: Appl. Phys.}\ }\textbf {\bibinfo {volume} {50}},\ \bibinfo {pages} {244001} (\bibinfo {year} {2017})}\BibitemShut {NoStop}%
\bibitem [{\citenamefont {Joannopoulos}\ \emph {et~al.}(2008)\citenamefont {Joannopoulos}, \citenamefont {Johnson}, \citenamefont {Winn},\ and\ \citenamefont {Meade}}]{Joannopoulos08}%
  \BibitemOpen
  \bibfield  {author} {\bibinfo {author} {\bibfnamefont {J.}~\bibnamefont {Joannopoulos}}, \bibinfo {author} {\bibfnamefont {S.~G.}\ \bibnamefont {Johnson}}, \bibinfo {author} {\bibfnamefont {J.~N.}\ \bibnamefont {Winn}},\ and\ \bibinfo {author} {\bibfnamefont {R.~D.}\ \bibnamefont {Meade}},\ }\href@noop {} {\emph {\bibinfo {title} {Photonic Crystals: Molding the Flow of Light}}}\ (\bibinfo  {publisher} {Princeton University Press},\ \bibinfo {year} {2008})\BibitemShut {NoStop}%
\bibitem [{\citenamefont {Tacchi}\ \emph {et~al.}(2012)\citenamefont {Tacchi}, \citenamefont {Duerr}, \citenamefont {Klos}, \citenamefont {Madami}, \citenamefont {Neusser}, \citenamefont {Gubbiotti}, \citenamefont {Carlotti}, \citenamefont {Krawczyk},\ and\ \citenamefont {Grundler}}]{Tacchi12}%
  \BibitemOpen
  \bibfield  {author} {\bibinfo {author} {\bibfnamefont {S.}~\bibnamefont {Tacchi}}, \bibinfo {author} {\bibfnamefont {G.}~\bibnamefont {Duerr}}, \bibinfo {author} {\bibfnamefont {J.~W.}\ \bibnamefont {Klos}}, \bibinfo {author} {\bibfnamefont {M.}~\bibnamefont {Madami}}, \bibinfo {author} {\bibfnamefont {S.}~\bibnamefont {Neusser}}, \bibinfo {author} {\bibfnamefont {G.}~\bibnamefont {Gubbiotti}}, \bibinfo {author} {\bibfnamefont {G.}~\bibnamefont {Carlotti}}, \bibinfo {author} {\bibfnamefont {M.}~\bibnamefont {Krawczyk}},\ and\ \bibinfo {author} {\bibfnamefont {D.}~\bibnamefont {Grundler}},\ }\href {https://doi.org/10.1103/PhysRevLett.109.137202} {\bibfield  {journal} {\bibinfo  {journal} {Phys. Rev. Lett.}\ }\textbf {\bibinfo {volume} {109}},\ \bibinfo {pages} {137202} (\bibinfo {year} {2012})}\BibitemShut {NoStop}%
\bibitem [{\citenamefont {Kruglyak}\ \emph {et~al.}(2010)\citenamefont {Kruglyak}, \citenamefont {Keatley}, \citenamefont {Neudert}, \citenamefont {Hicken}, \citenamefont {Childress},\ and\ \citenamefont {Katine}}]{Kruglyak2010}%
  \BibitemOpen
  \bibfield  {author} {\bibinfo {author} {\bibfnamefont {V.~V.}\ \bibnamefont {Kruglyak}}, \bibinfo {author} {\bibfnamefont {P.~S.}\ \bibnamefont {Keatley}}, \bibinfo {author} {\bibfnamefont {A.}~\bibnamefont {Neudert}}, \bibinfo {author} {\bibfnamefont {R.~J.}\ \bibnamefont {Hicken}}, \bibinfo {author} {\bibfnamefont {J.~R.}\ \bibnamefont {Childress}},\ and\ \bibinfo {author} {\bibfnamefont {J.~A.}\ \bibnamefont {Katine}},\ }\href {https://doi.org/10.1103/PhysRevLett.104.027201} {\bibfield  {journal} {\bibinfo  {journal} {Phys. Rev. Lett.}\ }\textbf {\bibinfo {volume} {104}},\ \bibinfo {pages} {027201} (\bibinfo {year} {2010})}\BibitemShut {NoStop}%
\bibitem [{\citenamefont {Tacchi}\ \emph {et~al.}(2011)\citenamefont {Tacchi}, \citenamefont {Montoncello}, \citenamefont {Madami}, \citenamefont {Gubbiotti}, \citenamefont {Carlotti}, \citenamefont {Giovannini}, \citenamefont {Zivieri}, \citenamefont {Nizzoli}, \citenamefont {Jain}, \citenamefont {Adeyeye},\ and\ \citenamefont {Singh}}]{Tacchi2011}%
  \BibitemOpen
  \bibfield  {author} {\bibinfo {author} {\bibfnamefont {S.}~\bibnamefont {Tacchi}}, \bibinfo {author} {\bibfnamefont {F.}~\bibnamefont {Montoncello}}, \bibinfo {author} {\bibfnamefont {M.}~\bibnamefont {Madami}}, \bibinfo {author} {\bibfnamefont {G.}~\bibnamefont {Gubbiotti}}, \bibinfo {author} {\bibfnamefont {G.}~\bibnamefont {Carlotti}}, \bibinfo {author} {\bibfnamefont {L.}~\bibnamefont {Giovannini}}, \bibinfo {author} {\bibfnamefont {R.}~\bibnamefont {Zivieri}}, \bibinfo {author} {\bibfnamefont {F.}~\bibnamefont {Nizzoli}}, \bibinfo {author} {\bibfnamefont {S.}~\bibnamefont {Jain}}, \bibinfo {author} {\bibfnamefont {a.~O.}\ \bibnamefont {Adeyeye}},\ and\ \bibinfo {author} {\bibfnamefont {N.}~\bibnamefont {Singh}},\ }\href {https://doi.org/10.1103/PhysRevLett.107.127204} {\bibfield  {journal} {\bibinfo  {journal} {Phys. Rev. Lett.}\ }\textbf {\bibinfo {volume} {107}},\ \bibinfo {pages} {127204} (\bibinfo {year} {2011})}\BibitemShut {NoStop}%
\bibitem [{\citenamefont {Saha}\ \emph {et~al.}(2013)\citenamefont {Saha}, \citenamefont {Mandal}, \citenamefont {Barman}, \citenamefont {Kumar}, \citenamefont {Rana}, \citenamefont {Fukuma}, \citenamefont {Sugimoto}, \citenamefont {Otani},\ and\ \citenamefont {Barman}}]{Saha2013}%
  \BibitemOpen
  \bibfield  {author} {\bibinfo {author} {\bibfnamefont {S.}~\bibnamefont {Saha}}, \bibinfo {author} {\bibfnamefont {R.}~\bibnamefont {Mandal}}, \bibinfo {author} {\bibfnamefont {S.}~\bibnamefont {Barman}}, \bibinfo {author} {\bibfnamefont {D.}~\bibnamefont {Kumar}}, \bibinfo {author} {\bibfnamefont {B.}~\bibnamefont {Rana}}, \bibinfo {author} {\bibfnamefont {Y.}~\bibnamefont {Fukuma}}, \bibinfo {author} {\bibfnamefont {S.}~\bibnamefont {Sugimoto}}, \bibinfo {author} {\bibfnamefont {Y.}~\bibnamefont {Otani}},\ and\ \bibinfo {author} {\bibfnamefont {A.}~\bibnamefont {Barman}},\ }\href {https://doi.org/10.1002/adfm.201202545} {\bibfield  {journal} {\bibinfo  {journal} {Adv. Funct. Mater.}\ }\textbf {\bibinfo {volume} {23}},\ \bibinfo {pages} {2378} (\bibinfo {year} {2013})}\BibitemShut {NoStop}%
\bibitem [{\citenamefont {Kostylev}\ \emph {et~al.}(2008)\citenamefont {Kostylev}, \citenamefont {Gubbiotti}, \citenamefont {Carlotti}, \citenamefont {Socino}, \citenamefont {Tacchi}, \citenamefont {Wang}, \citenamefont {Singh}, \citenamefont {Adeyeye},\ and\ \citenamefont {Stamps}}]{Kostylev2008}%
  \BibitemOpen
  \bibfield  {author} {\bibinfo {author} {\bibfnamefont {M.}~\bibnamefont {Kostylev}}, \bibinfo {author} {\bibfnamefont {G.}~\bibnamefont {Gubbiotti}}, \bibinfo {author} {\bibfnamefont {G.}~\bibnamefont {Carlotti}}, \bibinfo {author} {\bibfnamefont {G.}~\bibnamefont {Socino}}, \bibinfo {author} {\bibfnamefont {S.}~\bibnamefont {Tacchi}}, \bibinfo {author} {\bibfnamefont {C.}~\bibnamefont {Wang}}, \bibinfo {author} {\bibfnamefont {N.}~\bibnamefont {Singh}}, \bibinfo {author} {\bibfnamefont {A.~O.}\ \bibnamefont {Adeyeye}},\ and\ \bibinfo {author} {\bibfnamefont {R.~L.}\ \bibnamefont {Stamps}},\ }\href {https://doi.org/10.1063/1.2831792} {\bibfield  {journal} {\bibinfo  {journal} {J. Appl. Phys.}\ }\textbf {\bibinfo {volume} {103}},\ \bibinfo {pages} {07C507} (\bibinfo {year} {2008})}\BibitemShut {NoStop}%
\bibitem [{\citenamefont {Mandal}\ \emph {et~al.}(2013)\citenamefont {Mandal}, \citenamefont {Laha}, \citenamefont {Das}, \citenamefont {Saha}, \citenamefont {Barman}, \citenamefont {Raychaudhuri},\ and\ \citenamefont {Barman}}]{Mandal2013}%
  \BibitemOpen
  \bibfield  {author} {\bibinfo {author} {\bibfnamefont {R.}~\bibnamefont {Mandal}}, \bibinfo {author} {\bibfnamefont {P.}~\bibnamefont {Laha}}, \bibinfo {author} {\bibfnamefont {K.}~\bibnamefont {Das}}, \bibinfo {author} {\bibfnamefont {S.}~\bibnamefont {Saha}}, \bibinfo {author} {\bibfnamefont {S.}~\bibnamefont {Barman}}, \bibinfo {author} {\bibfnamefont {A.~K.}\ \bibnamefont {Raychaudhuri}},\ and\ \bibinfo {author} {\bibfnamefont {A.}~\bibnamefont {Barman}},\ }\href {https://doi.org/10.1063/1.4860959} {\bibfield  {journal} {\bibinfo  {journal} {Appl. Phys. Lett.}\ }\textbf {\bibinfo {volume} {103}},\ \bibinfo {pages} {262410} (\bibinfo {year} {2013})}\BibitemShut {NoStop}%
\bibitem [{\citenamefont {Martin}\ \emph {et~al.}(2003)\citenamefont {Martin}, \citenamefont {Nogu{\'{e}}s}, \citenamefont {Liu}, \citenamefont {Vicent},\ and\ \citenamefont {Schuller}}]{Martin2003}%
  \BibitemOpen
  \bibfield  {author} {\bibinfo {author} {\bibfnamefont {J.}~\bibnamefont {Martin}}, \bibinfo {author} {\bibfnamefont {J.}~\bibnamefont {Nogu{\'{e}}s}}, \bibinfo {author} {\bibfnamefont {K.}~\bibnamefont {Liu}}, \bibinfo {author} {\bibfnamefont {J.}~\bibnamefont {Vicent}},\ and\ \bibinfo {author} {\bibfnamefont {I.~K.}\ \bibnamefont {Schuller}},\ }\href {https://doi.org/10.1016/S0304-8853(02)00898-3} {\bibfield  {journal} {\bibinfo  {journal} {J. Magn. Magn. Mater.}\ }\textbf {\bibinfo {volume} {256}},\ \bibinfo {pages} {449} (\bibinfo {year} {2003})}\BibitemShut {NoStop}%
\bibitem [{\citenamefont {Lau}\ and\ \citenamefont {Shaw}(2011)}]{Lau2011}%
  \BibitemOpen
  \bibfield  {author} {\bibinfo {author} {\bibfnamefont {J.~W.}\ \bibnamefont {Lau}}\ and\ \bibinfo {author} {\bibfnamefont {J.~M.}\ \bibnamefont {Shaw}},\ }\href {https://doi.org/10.1088/0022-3727/44/30/303001} {\bibfield  {journal} {\bibinfo  {journal} {J. Phys. D: Appl. Phys.}\ }\textbf {\bibinfo {volume} {44}},\ \bibinfo {pages} {303001} (\bibinfo {year} {2011})}\BibitemShut {NoStop}%
\bibitem [{\citenamefont {Carter}\ \emph {et~al.}(1982)\citenamefont {Carter}, \citenamefont {Owens}, \citenamefont {Smith},\ and\ \citenamefont {Reed}}]{Carter1982}%
  \BibitemOpen
  \bibfield  {author} {\bibinfo {author} {\bibfnamefont {R.~L.}\ \bibnamefont {Carter}}, \bibinfo {author} {\bibfnamefont {J.~M.}\ \bibnamefont {Owens}}, \bibinfo {author} {\bibfnamefont {C.~V.}\ \bibnamefont {Smith}},\ and\ \bibinfo {author} {\bibfnamefont {K.~W.}\ \bibnamefont {Reed}},\ }\href {https://doi.org/10.1063/1.330928} {\bibfield  {journal} {\bibinfo  {journal} {J. Appl. Phys.}\ }\textbf {\bibinfo {volume} {53}},\ \bibinfo {pages} {2655} (\bibinfo {year} {1982})}\BibitemShut {NoStop}%
\bibitem [{\citenamefont {Urb{\'{a}}nek}\ \emph {et~al.}(2018)\citenamefont {Urb{\'{a}}nek}, \citenamefont {Flaj{\v{s}}man}, \citenamefont {K{\v{r}}i{\v{z}}{\'{a}}kov{\'{a}}}, \citenamefont {Gloss}, \citenamefont {Hork{\'{y}}}, \citenamefont {Schmid},\ and\ \citenamefont {Varga}}]{Urbanek2018}%
  \BibitemOpen
  \bibfield  {author} {\bibinfo {author} {\bibfnamefont {M.}~\bibnamefont {Urb{\'{a}}nek}}, \bibinfo {author} {\bibfnamefont {L.}~\bibnamefont {Flaj{\v{s}}man}}, \bibinfo {author} {\bibfnamefont {V.}~\bibnamefont {K{\v{r}}i{\v{z}}{\'{a}}kov{\'{a}}}}, \bibinfo {author} {\bibfnamefont {J.}~\bibnamefont {Gloss}}, \bibinfo {author} {\bibfnamefont {M.}~\bibnamefont {Hork{\'{y}}}}, \bibinfo {author} {\bibfnamefont {M.}~\bibnamefont {Schmid}},\ and\ \bibinfo {author} {\bibfnamefont {P.}~\bibnamefont {Varga}},\ }\href {https://doi.org/10.1063/1.5029367} {\bibfield  {journal} {\bibinfo  {journal} {APL Mater.}\ }\textbf {\bibinfo {volume} {6}},\ \bibinfo {pages} {060701} (\bibinfo {year} {2018})}\BibitemShut {NoStop}%
\bibitem [{\citenamefont {Fassbender}\ \emph {et~al.}(2009)\citenamefont {Fassbender}, \citenamefont {Strache}, \citenamefont {Liedke}, \citenamefont {Mark{\'{o}}}, \citenamefont {Wintz}, \citenamefont {Lenz}, \citenamefont {Keller}, \citenamefont {Facsko}, \citenamefont {M{\"{o}}nch},\ and\ \citenamefont {McCord}}]{Fassbender2009}%
  \BibitemOpen
  \bibfield  {author} {\bibinfo {author} {\bibfnamefont {J.}~\bibnamefont {Fassbender}}, \bibinfo {author} {\bibfnamefont {T.}~\bibnamefont {Strache}}, \bibinfo {author} {\bibfnamefont {M.~O.}\ \bibnamefont {Liedke}}, \bibinfo {author} {\bibfnamefont {D.}~\bibnamefont {Mark{\'{o}}}}, \bibinfo {author} {\bibfnamefont {S.}~\bibnamefont {Wintz}}, \bibinfo {author} {\bibfnamefont {K.}~\bibnamefont {Lenz}}, \bibinfo {author} {\bibfnamefont {A.}~\bibnamefont {Keller}}, \bibinfo {author} {\bibfnamefont {S.}~\bibnamefont {Facsko}}, \bibinfo {author} {\bibfnamefont {I.}~\bibnamefont {M{\"{o}}nch}},\ and\ \bibinfo {author} {\bibfnamefont {J.}~\bibnamefont {McCord}},\ }\href {https://doi.org/10.1088/1367-2630/11/12/125002} {\bibfield  {journal} {\bibinfo  {journal} {New J. Phys.}\ }\textbf {\bibinfo {volume} {11}},\ \bibinfo {pages} {125002} (\bibinfo {year} {2009})}\BibitemShut {NoStop}%
\bibitem [{\citenamefont {Obry}\ \emph {et~al.}(2013)\citenamefont {Obry}, \citenamefont {Pirro}, \citenamefont {Br{\"{a}}cher}, \citenamefont {Chumak}, \citenamefont {Osten}, \citenamefont {Ciubotaru}, \citenamefont {Serga}, \citenamefont {Fassbender},\ and\ \citenamefont {Hillebrands}}]{Obry2013}%
  \BibitemOpen
  \bibfield  {author} {\bibinfo {author} {\bibfnamefont {B.}~\bibnamefont {Obry}}, \bibinfo {author} {\bibfnamefont {P.}~\bibnamefont {Pirro}}, \bibinfo {author} {\bibfnamefont {T.}~\bibnamefont {Br{\"{a}}cher}}, \bibinfo {author} {\bibfnamefont {A.~V.}\ \bibnamefont {Chumak}}, \bibinfo {author} {\bibfnamefont {J.}~\bibnamefont {Osten}}, \bibinfo {author} {\bibfnamefont {F.}~\bibnamefont {Ciubotaru}}, \bibinfo {author} {\bibfnamefont {A.~A.}\ \bibnamefont {Serga}}, \bibinfo {author} {\bibfnamefont {J.}~\bibnamefont {Fassbender}},\ and\ \bibinfo {author} {\bibfnamefont {B.}~\bibnamefont {Hillebrands}},\ }\href {https://doi.org/10.1063/1.4807721} {\bibfield  {journal} {\bibinfo  {journal} {Appl. Phys. Lett.}\ }\textbf {\bibinfo {volume} {102}},\ \bibinfo {pages} {202403} (\bibinfo {year} {2013})}\BibitemShut {NoStop}%
\bibitem [{\citenamefont {Wawro}\ \emph {et~al.}(2018)\citenamefont {Wawro}, \citenamefont {Kurant}, \citenamefont {Jakubowski}, \citenamefont {Tekielak}, \citenamefont {Pietruczik}, \citenamefont {B{\"{o}}ttger},\ and\ \citenamefont {Maziewski}}]{Wawro2018}%
  \BibitemOpen
  \bibfield  {author} {\bibinfo {author} {\bibfnamefont {A.}~\bibnamefont {Wawro}}, \bibinfo {author} {\bibfnamefont {Z.}~\bibnamefont {Kurant}}, \bibinfo {author} {\bibfnamefont {M.}~\bibnamefont {Jakubowski}}, \bibinfo {author} {\bibfnamefont {M.}~\bibnamefont {Tekielak}}, \bibinfo {author} {\bibfnamefont {A.}~\bibnamefont {Pietruczik}}, \bibinfo {author} {\bibfnamefont {R.}~\bibnamefont {B{\"{o}}ttger}},\ and\ \bibinfo {author} {\bibfnamefont {A.}~\bibnamefont {Maziewski}},\ }\href {https://doi.org/10.1103/PhysRevApplied.9.014029} {\bibfield  {journal} {\bibinfo  {journal} {Phys. Rev. Appl.}\ }\textbf {\bibinfo {volume} {9}},\ \bibinfo {pages} {014029} (\bibinfo {year} {2018})}\BibitemShut {NoStop}%
\bibitem [{\citenamefont {Yu}\ \emph {et~al.}(2021)\citenamefont {Yu}, \citenamefont {Xiao},\ and\ \citenamefont {Schultheiss}}]{Yu2021}%
  \BibitemOpen
  \bibfield  {author} {\bibinfo {author} {\bibfnamefont {H.}~\bibnamefont {Yu}}, \bibinfo {author} {\bibfnamefont {J.}~\bibnamefont {Xiao}},\ and\ \bibinfo {author} {\bibfnamefont {H.}~\bibnamefont {Schultheiss}},\ }\href {https://doi.org/10.1016/j.physrep.2020.12.004} {\bibfield  {journal} {\bibinfo  {journal} {Phys. Rep.}\ }\textbf {\bibinfo {volume} {905}},\ \bibinfo {pages} {1} (\bibinfo {year} {2021})}\BibitemShut {NoStop}%
\bibitem [{\citenamefont {Banerjee}\ \emph {et~al.}(2017)\citenamefont {Banerjee}, \citenamefont {Gruszecki}, \citenamefont {Klos}, \citenamefont {Hellwig}, \citenamefont {Krawczyk},\ and\ \citenamefont {Barman}}]{Banerjee2017}%
  \BibitemOpen
  \bibfield  {author} {\bibinfo {author} {\bibfnamefont {C.}~\bibnamefont {Banerjee}}, \bibinfo {author} {\bibfnamefont {P.}~\bibnamefont {Gruszecki}}, \bibinfo {author} {\bibfnamefont {J.~W.}\ \bibnamefont {Klos}}, \bibinfo {author} {\bibfnamefont {O.}~\bibnamefont {Hellwig}}, \bibinfo {author} {\bibfnamefont {M.}~\bibnamefont {Krawczyk}},\ and\ \bibinfo {author} {\bibfnamefont {A.}~\bibnamefont {Barman}},\ }\href {https://doi.org/10.1103/PHYSREVB.96.024421/FIGURES/3/MEDIUM} {\bibfield  {journal} {\bibinfo  {journal} {Phys. Rev. B}\ }\textbf {\bibinfo {volume} {96}},\ \bibinfo {pages} {024421} (\bibinfo {year} {2017})}\BibitemShut {NoStop}%
\bibitem [{\citenamefont {Szulc}\ \emph {et~al.}(2022)\citenamefont {Szulc}, \citenamefont {Tacchi}, \citenamefont {Hierro-Rodr{\'{i}}guez}, \citenamefont {D{\'{i}}az}, \citenamefont {Gruszecki}, \citenamefont {Graczyk}, \citenamefont {Quir{\'{o}}s}, \citenamefont {Mark{\'{o}}}, \citenamefont {Mart{\'{i}}n}, \citenamefont {V{\'{e}}lez}, \citenamefont {Schmool}, \citenamefont {Carlotti}, \citenamefont {Krawczyk},\ and\ \citenamefont {{\'{A}}lvarez-Prado}}]{Szulc2022}%
  \BibitemOpen
  \bibfield  {author} {\bibinfo {author} {\bibfnamefont {K.}~\bibnamefont {Szulc}}, \bibinfo {author} {\bibfnamefont {S.}~\bibnamefont {Tacchi}}, \bibinfo {author} {\bibfnamefont {A.}~\bibnamefont {Hierro-Rodr{\'{i}}guez}}, \bibinfo {author} {\bibfnamefont {J.}~\bibnamefont {D{\'{i}}az}}, \bibinfo {author} {\bibfnamefont {P.}~\bibnamefont {Gruszecki}}, \bibinfo {author} {\bibfnamefont {P.}~\bibnamefont {Graczyk}}, \bibinfo {author} {\bibfnamefont {C.}~\bibnamefont {Quir{\'{o}}s}}, \bibinfo {author} {\bibfnamefont {D.}~\bibnamefont {Mark{\'{o}}}}, \bibinfo {author} {\bibfnamefont {J.~I.}\ \bibnamefont {Mart{\'{i}}n}}, \bibinfo {author} {\bibfnamefont {M.}~\bibnamefont {V{\'{e}}lez}}, \bibinfo {author} {\bibfnamefont {D.~S.}\ \bibnamefont {Schmool}}, \bibinfo {author} {\bibfnamefont {G.}~\bibnamefont {Carlotti}}, \bibinfo {author} {\bibfnamefont {M.}~\bibnamefont {Krawczyk}},\ and\ \bibinfo {author} {\bibfnamefont {L.~M.}\ \bibnamefont {{\'{A}}lvarez-Prado}},\ }\href {https://doi.org/10.1021/acsnano.2c04256}
  {\bibfield  {journal} {\bibinfo  {journal} {ACS Nano}\ }\textbf {\bibinfo {volume} {16}},\ \bibinfo {pages} {14168} (\bibinfo {year} {2022})}\BibitemShut {NoStop}%
\bibitem [{\citenamefont {Di{\'{a}}z}\ \emph {et~al.}(2020)\citenamefont {Di{\'{a}}z}, \citenamefont {Hirosawa}, \citenamefont {Klinovaja},\ and\ \citenamefont {Loss}}]{Diaz2020}%
  \BibitemOpen
  \bibfield  {author} {\bibinfo {author} {\bibfnamefont {S.~A.}\ \bibnamefont {Di{\'{a}}z}}, \bibinfo {author} {\bibfnamefont {T.}~\bibnamefont {Hirosawa}}, \bibinfo {author} {\bibfnamefont {J.}~\bibnamefont {Klinovaja}},\ and\ \bibinfo {author} {\bibfnamefont {D.}~\bibnamefont {Loss}},\ }\href {https://doi.org/10.1103/PHYSREVRESEARCH.2.013231} {\bibfield  {journal} {\bibinfo  {journal} {Phys. Rev. Res.}\ }\textbf {\bibinfo {volume} {2}},\ \bibinfo {pages} {013231} (\bibinfo {year} {2020})}\BibitemShut {NoStop}%
\bibitem [{\citenamefont {Chen}\ and\ \citenamefont {Ma}(2021)}]{Chen2021}%
  \BibitemOpen
  \bibfield  {author} {\bibinfo {author} {\bibfnamefont {Z.}~\bibnamefont {Chen}}\ and\ \bibinfo {author} {\bibfnamefont {F.}~\bibnamefont {Ma}},\ }\href {https://doi.org/10.1063/5.0061832} {\bibfield  {journal} {\bibinfo  {journal} {J. Appl. Phys.}\ }\textbf {\bibinfo {volume} {130}},\ \bibinfo {pages} {090901} (\bibinfo {year} {2021})}\BibitemShut {NoStop}%
\bibitem [{\citenamefont {Takagi}\ \emph {et~al.}(2021)\citenamefont {Takagi}, \citenamefont {Garst}, \citenamefont {Sahliger}, \citenamefont {Back}, \citenamefont {Tokura},\ and\ \citenamefont {Seki}}]{Takagi2021}%
  \BibitemOpen
  \bibfield  {author} {\bibinfo {author} {\bibfnamefont {R.}~\bibnamefont {Takagi}}, \bibinfo {author} {\bibfnamefont {M.}~\bibnamefont {Garst}}, \bibinfo {author} {\bibfnamefont {J.}~\bibnamefont {Sahliger}}, \bibinfo {author} {\bibfnamefont {C.~H.}\ \bibnamefont {Back}}, \bibinfo {author} {\bibfnamefont {Y.}~\bibnamefont {Tokura}},\ and\ \bibinfo {author} {\bibfnamefont {S.}~\bibnamefont {Seki}},\ }\href {https://doi.org/10.1103/PHYSREVB.104.144410} {\bibfield  {journal} {\bibinfo  {journal} {Phys. Rev. B}\ }\textbf {\bibinfo {volume} {104}},\ \bibinfo {pages} {144410} (\bibinfo {year} {2021})}\BibitemShut {NoStop}%
\bibitem [{\citenamefont {Bassotti}\ \emph {et~al.}(2022)\citenamefont {Bassotti}, \citenamefont {Silvani},\ and\ \citenamefont {Carlotti}}]{Bassotti2022}%
  \BibitemOpen
  \bibfield  {author} {\bibinfo {author} {\bibfnamefont {M.}~\bibnamefont {Bassotti}}, \bibinfo {author} {\bibfnamefont {R.}~\bibnamefont {Silvani}},\ and\ \bibinfo {author} {\bibfnamefont {G.}~\bibnamefont {Carlotti}},\ }\href {https://doi.org/10.1109/LMAG.2021.3136152} {\bibfield  {journal} {\bibinfo  {journal} {IEEE Magn. Lett.}\ }\textbf {\bibinfo {volume} {13}},\ \bibinfo {pages} {6101505} (\bibinfo {year} {2022})}\BibitemShut {NoStop}%
\bibitem [{\citenamefont {Chen}\ \emph {et~al.}(2018)\citenamefont {Chen}, \citenamefont {Liu}, \citenamefont {Liu}, \citenamefont {Xiao}, \citenamefont {Xia}, \citenamefont {Bauer}, \citenamefont {Wu},\ and\ \citenamefont {Yu}}]{chen2018strong}%
  \BibitemOpen
  \bibfield  {author} {\bibinfo {author} {\bibfnamefont {J.}~\bibnamefont {Chen}}, \bibinfo {author} {\bibfnamefont {C.}~\bibnamefont {Liu}}, \bibinfo {author} {\bibfnamefont {T.}~\bibnamefont {Liu}}, \bibinfo {author} {\bibfnamefont {Y.}~\bibnamefont {Xiao}}, \bibinfo {author} {\bibfnamefont {K.}~\bibnamefont {Xia}}, \bibinfo {author} {\bibfnamefont {G.~E.}\ \bibnamefont {Bauer}}, \bibinfo {author} {\bibfnamefont {M.}~\bibnamefont {Wu}},\ and\ \bibinfo {author} {\bibfnamefont {H.}~\bibnamefont {Yu}},\ }\href@noop {} {\bibfield  {journal} {\bibinfo  {journal} {Phys. Rev. Lett.}\ }\textbf {\bibinfo {volume} {120}},\ \bibinfo {pages} {217202} (\bibinfo {year} {2018})}\BibitemShut {NoStop}%
\bibitem [{\citenamefont {Sud}\ \emph {et~al.}(2020)\citenamefont {Sud}, \citenamefont {Zollitsch}, \citenamefont {Kamimaki}, \citenamefont {Dion}, \citenamefont {Khan}, \citenamefont {Iihama}, \citenamefont {Mizukami},\ and\ \citenamefont {Kurebayashi}}]{sud2020tunable}%
  \BibitemOpen
  \bibfield  {author} {\bibinfo {author} {\bibfnamefont {A.}~\bibnamefont {Sud}}, \bibinfo {author} {\bibfnamefont {C.}~\bibnamefont {Zollitsch}}, \bibinfo {author} {\bibfnamefont {A.}~\bibnamefont {Kamimaki}}, \bibinfo {author} {\bibfnamefont {T.}~\bibnamefont {Dion}}, \bibinfo {author} {\bibfnamefont {S.}~\bibnamefont {Khan}}, \bibinfo {author} {\bibfnamefont {S.}~\bibnamefont {Iihama}}, \bibinfo {author} {\bibfnamefont {S.}~\bibnamefont {Mizukami}},\ and\ \bibinfo {author} {\bibfnamefont {H.}~\bibnamefont {Kurebayashi}},\ }\href@noop {} {\bibfield  {journal} {\bibinfo  {journal} {Phys. Rev. B}\ }\textbf {\bibinfo {volume} {102}},\ \bibinfo {pages} {100403} (\bibinfo {year} {2020})}\BibitemShut {NoStop}%
\bibitem [{\citenamefont {Shiota}\ \emph {et~al.}(2020)\citenamefont {Shiota}, \citenamefont {Taniguchi}, \citenamefont {Ishibashi}, \citenamefont {Moriyama},\ and\ \citenamefont {Ono}}]{shiota2020tunable}%
  \BibitemOpen
  \bibfield  {author} {\bibinfo {author} {\bibfnamefont {Y.}~\bibnamefont {Shiota}}, \bibinfo {author} {\bibfnamefont {T.}~\bibnamefont {Taniguchi}}, \bibinfo {author} {\bibfnamefont {M.}~\bibnamefont {Ishibashi}}, \bibinfo {author} {\bibfnamefont {T.}~\bibnamefont {Moriyama}},\ and\ \bibinfo {author} {\bibfnamefont {T.}~\bibnamefont {Ono}},\ }\href@noop {} {\bibfield  {journal} {\bibinfo  {journal} {Phys. Rev. Lett.}\ }\textbf {\bibinfo {volume} {125}},\ \bibinfo {pages} {017203} (\bibinfo {year} {2020})}\BibitemShut {NoStop}%
\bibitem [{\citenamefont {Li}\ \emph {et~al.}(2020)\citenamefont {Li}, \citenamefont {Zhang}, \citenamefont {Tyberkevych}, \citenamefont {Kwok}, \citenamefont {Hoffmann},\ and\ \citenamefont {Novosad}}]{li2020hybrid}%
  \BibitemOpen
  \bibfield  {author} {\bibinfo {author} {\bibfnamefont {Y.}~\bibnamefont {Li}}, \bibinfo {author} {\bibfnamefont {W.}~\bibnamefont {Zhang}}, \bibinfo {author} {\bibfnamefont {V.}~\bibnamefont {Tyberkevych}}, \bibinfo {author} {\bibfnamefont {W.-K.}\ \bibnamefont {Kwok}}, \bibinfo {author} {\bibfnamefont {A.}~\bibnamefont {Hoffmann}},\ and\ \bibinfo {author} {\bibfnamefont {V.}~\bibnamefont {Novosad}},\ }\href@noop {} {\bibfield  {journal} {\bibinfo  {journal} {J. Appl. Phys.}\ }\textbf {\bibinfo {volume} {128}},\ \bibinfo {pages} {130902} (\bibinfo {year} {2020})}\BibitemShut {NoStop}%
\bibitem [{\citenamefont {Dai}\ \emph {et~al.}(2020)\citenamefont {Dai}, \citenamefont {Xie}, \citenamefont {Pan},\ and\ \citenamefont {Ma}}]{dai2020strong}%
  \BibitemOpen
  \bibfield  {author} {\bibinfo {author} {\bibfnamefont {C.}~\bibnamefont {Dai}}, \bibinfo {author} {\bibfnamefont {K.}~\bibnamefont {Xie}}, \bibinfo {author} {\bibfnamefont {Z.}~\bibnamefont {Pan}},\ and\ \bibinfo {author} {\bibfnamefont {F.}~\bibnamefont {Ma}},\ }\href@noop {} {\bibfield  {journal} {\bibinfo  {journal} {J. Appl. Phys.}\ }\textbf {\bibinfo {volume} {127}},\ \bibinfo {pages} {203902} (\bibinfo {year} {2020})}\BibitemShut {NoStop}%
\bibitem [{\citenamefont {Li}\ \emph {et~al.}(2022)\citenamefont {Li}, \citenamefont {Ma}, \citenamefont {Chen}, \citenamefont {Xie},\ and\ \citenamefont {Ma}}]{li2022interaction}%
  \BibitemOpen
  \bibfield  {author} {\bibinfo {author} {\bibfnamefont {Z.}~\bibnamefont {Li}}, \bibinfo {author} {\bibfnamefont {M.}~\bibnamefont {Ma}}, \bibinfo {author} {\bibfnamefont {Z.}~\bibnamefont {Chen}}, \bibinfo {author} {\bibfnamefont {K.}~\bibnamefont {Xie}},\ and\ \bibinfo {author} {\bibfnamefont {F.}~\bibnamefont {Ma}},\ }\href@noop {} {\bibfield  {journal} {\bibinfo  {journal} {J. Appl. Phys.}\ }\textbf {\bibinfo {volume} {132}},\ \bibinfo {pages} {210702} (\bibinfo {year} {2022})}\BibitemShut {NoStop}%
\bibitem [{\citenamefont {Pal}\ \emph {et~al.}(2014)\citenamefont {Pal}, \citenamefont {Klos}, \citenamefont {Das}, \citenamefont {Hellwig}, \citenamefont {Gruszecki}, \citenamefont {Krawczyk},\ and\ \citenamefont {Barman}}]{Pal2014}%
  \BibitemOpen
  \bibfield  {author} {\bibinfo {author} {\bibfnamefont {S.}~\bibnamefont {Pal}}, \bibinfo {author} {\bibfnamefont {J.~W.}\ \bibnamefont {Klos}}, \bibinfo {author} {\bibfnamefont {K.}~\bibnamefont {Das}}, \bibinfo {author} {\bibfnamefont {O.}~\bibnamefont {Hellwig}}, \bibinfo {author} {\bibfnamefont {P.}~\bibnamefont {Gruszecki}}, \bibinfo {author} {\bibfnamefont {M.}~\bibnamefont {Krawczyk}},\ and\ \bibinfo {author} {\bibfnamefont {A.}~\bibnamefont {Barman}},\ }\href {https://doi.org/10.1063/1.4898774} {\bibfield  {journal} {\bibinfo  {journal} {Appl. Phys. Lett.}\ }\textbf {\bibinfo {volume} {105}},\ \bibinfo {pages} {162408} (\bibinfo {year} {2014})}\BibitemShut {NoStop}%
\bibitem [{\citenamefont {Pan}\ \emph {et~al.}(2020)\citenamefont {Pan}, \citenamefont {Mondal}, \citenamefont {Zelent}, \citenamefont {Szwierz}, \citenamefont {Pal}, \citenamefont {Hellwig}, \citenamefont {Krawczyk},\ and\ \citenamefont {Barman}}]{Pan2020c}%
  \BibitemOpen
  \bibfield  {author} {\bibinfo {author} {\bibfnamefont {S.}~\bibnamefont {Pan}}, \bibinfo {author} {\bibfnamefont {S.}~\bibnamefont {Mondal}}, \bibinfo {author} {\bibfnamefont {M.}~\bibnamefont {Zelent}}, \bibinfo {author} {\bibfnamefont {R.}~\bibnamefont {Szwierz}}, \bibinfo {author} {\bibfnamefont {S.}~\bibnamefont {Pal}}, \bibinfo {author} {\bibfnamefont {O.}~\bibnamefont {Hellwig}}, \bibinfo {author} {\bibfnamefont {M.}~\bibnamefont {Krawczyk}},\ and\ \bibinfo {author} {\bibfnamefont {A.}~\bibnamefont {Barman}},\ }\href {https://doi.org/10.1103/PhysRevB.101.014403} {\bibfield  {journal} {\bibinfo  {journal} {Phys. Rev. B}\ }\textbf {\bibinfo {volume} {101}},\ \bibinfo {pages} {014403} (\bibinfo {year} {2020})}\BibitemShut {NoStop}%
\bibitem [{\citenamefont {Mantion}\ and\ \citenamefont {Biziere}(2022)}]{Mantion2022}%
  \BibitemOpen
  \bibfield  {author} {\bibinfo {author} {\bibfnamefont {S.}~\bibnamefont {Mantion}}\ and\ \bibinfo {author} {\bibfnamefont {N.}~\bibnamefont {Biziere}},\ }\href {https://doi.org/10.1063/5.0085623} {\bibfield  {journal} {\bibinfo  {journal} {J. Appl. Phys.}\ }\textbf {\bibinfo {volume} {131}},\ \bibinfo {pages} {113905} (\bibinfo {year} {2022})}\BibitemShut {NoStop}%
\bibitem [{\citenamefont {Moalic}\ \emph {et~al.}(2022)\citenamefont {Moalic}, \citenamefont {Krawczyk},\ and\ \citenamefont {Zelent}}]{moalic2022spin}%
  \BibitemOpen
  \bibfield  {author} {\bibinfo {author} {\bibfnamefont {M.}~\bibnamefont {Moalic}}, \bibinfo {author} {\bibfnamefont {M.}~\bibnamefont {Krawczyk}},\ and\ \bibinfo {author} {\bibfnamefont {M.}~\bibnamefont {Zelent}},\ }\href@noop {} {\bibfield  {journal} {\bibinfo  {journal} {J. Appl. Phys.}\ }\textbf {\bibinfo {volume} {132}} (\bibinfo {year} {2022})}\BibitemShut {NoStop}%
\bibitem [{\citenamefont {Lemesh}\ and\ \citenamefont {Beach}(2018)}]{lemesh2018twisted}%
  \BibitemOpen
  \bibfield  {author} {\bibinfo {author} {\bibfnamefont {I.}~\bibnamefont {Lemesh}}\ and\ \bibinfo {author} {\bibfnamefont {G.~S.}\ \bibnamefont {Beach}},\ }\href@noop {} {\bibfield  {journal} {\bibinfo  {journal} {Phys. Rev. B}\ }\textbf {\bibinfo {volume} {98}},\ \bibinfo {pages} {104402} (\bibinfo {year} {2018})}\BibitemShut {NoStop}%
\bibitem [{\citenamefont {Vansteenkiste}\ \emph {et~al.}(2014)\citenamefont {Vansteenkiste}, \citenamefont {Leliaert}, \citenamefont {Dvornik}, \citenamefont {Helsen}, \citenamefont {Garcia-Sanchez},\ and\ \citenamefont {Van~Waeyenberge}}]{mumax_2014}%
  \BibitemOpen
  \bibfield  {author} {\bibinfo {author} {\bibfnamefont {A.}~\bibnamefont {Vansteenkiste}}, \bibinfo {author} {\bibfnamefont {J.}~\bibnamefont {Leliaert}}, \bibinfo {author} {\bibfnamefont {M.}~\bibnamefont {Dvornik}}, \bibinfo {author} {\bibfnamefont {M.}~\bibnamefont {Helsen}}, \bibinfo {author} {\bibfnamefont {F.}~\bibnamefont {Garcia-Sanchez}},\ and\ \bibinfo {author} {\bibfnamefont {B.}~\bibnamefont {Van~Waeyenberge}},\ }\href {https://doi.org/10.1063/1.4899186} {\bibfield  {journal} {\bibinfo  {journal} {AIP Adv.}\ }\textbf {\bibinfo {volume} {4}},\ \bibinfo {pages} {107133} (\bibinfo {year} {2014})}\BibitemShut {NoStop}%
\bibitem [{\citenamefont {Leliaert}\ \emph {et~al.}(2018)\citenamefont {Leliaert}, \citenamefont {Dvornik}, \citenamefont {Mulkers}, \citenamefont {De~Clercq}, \citenamefont {Milo{\v{s}}evi{\'{c}}},\ and\ \citenamefont {Van~Waeyenberge}}]{Leliaert2018}%
  \BibitemOpen
  \bibfield  {author} {\bibinfo {author} {\bibfnamefont {J.}~\bibnamefont {Leliaert}}, \bibinfo {author} {\bibfnamefont {M.}~\bibnamefont {Dvornik}}, \bibinfo {author} {\bibfnamefont {J.}~\bibnamefont {Mulkers}}, \bibinfo {author} {\bibfnamefont {J.}~\bibnamefont {De~Clercq}}, \bibinfo {author} {\bibfnamefont {M.~V.}\ \bibnamefont {Milo{\v{s}}evi{\'{c}}}},\ and\ \bibinfo {author} {\bibfnamefont {B.}~\bibnamefont {Van~Waeyenberge}},\ }\href {https://doi.org/10.1088/1361-6463/aaab1c} {\bibfield  {journal} {\bibinfo  {journal} {J. Phys. D: Appl. Phys.}\ }\textbf {\bibinfo {volume} {51}},\ \bibinfo {pages} {123002} (\bibinfo {year} {2018})}\BibitemShut {NoStop}%
\bibitem [{\citenamefont {Moalic}\ and\ \citenamefont {Zelent}(2023)}]{amumax2023}%
  \BibitemOpen
  \bibfield  {author} {\bibinfo {author} {\bibfnamefont {M.}~\bibnamefont {Moalic}}\ and\ \bibinfo {author} {\bibfnamefont {M.}~\bibnamefont {Zelent}},\ }\href {https://doi.org/10.5281/zenodo.10043142} {\bibinfo {title} {Mathieumoalic/amumax: 2023.10.26}} (\bibinfo {year} {2023})\BibitemShut {NoStop}%
\bibitem [{\citenamefont {Dugaev}\ \emph {et~al.}(2005)\citenamefont {Dugaev}, \citenamefont {Bruno}, \citenamefont {Canals},\ and\ \citenamefont {Lacroix}}]{Dugaev2005}%
  \BibitemOpen
  \bibfield  {author} {\bibinfo {author} {\bibfnamefont {V.~K.}\ \bibnamefont {Dugaev}}, \bibinfo {author} {\bibfnamefont {P.}~\bibnamefont {Bruno}}, \bibinfo {author} {\bibfnamefont {B.}~\bibnamefont {Canals}},\ and\ \bibinfo {author} {\bibfnamefont {C.}~\bibnamefont {Lacroix}},\ }\href {https://doi.org/10.1103/PhysRevB.72.024456} {\bibfield  {journal} {\bibinfo  {journal} {Phys. Rev. B}\ }\textbf {\bibinfo {volume} {72}},\ \bibinfo {pages} {024456} (\bibinfo {year} {2005})}\BibitemShut {NoStop}%
\bibitem [{\citenamefont {Varentcova}\ \emph {et~al.}(2020)\citenamefont {Varentcova}, \citenamefont {von Malottki}, \citenamefont {Potkina}, \citenamefont {Kwiatkowski}, \citenamefont {Heinze},\ and\ \citenamefont {Bessarab}}]{varentcova2020toward}%
  \BibitemOpen
  \bibfield  {author} {\bibinfo {author} {\bibfnamefont {A.~S.}\ \bibnamefont {Varentcova}}, \bibinfo {author} {\bibfnamefont {S.}~\bibnamefont {von Malottki}}, \bibinfo {author} {\bibfnamefont {M.~N.}\ \bibnamefont {Potkina}}, \bibinfo {author} {\bibfnamefont {G.}~\bibnamefont {Kwiatkowski}}, \bibinfo {author} {\bibfnamefont {S.}~\bibnamefont {Heinze}},\ and\ \bibinfo {author} {\bibfnamefont {P.~F.}\ \bibnamefont {Bessarab}},\ }\href@noop {} {\bibfield  {journal} {\bibinfo  {journal} {npj Computational Materials}\ }\textbf {\bibinfo {volume} {6}},\ \bibinfo {pages} {193} (\bibinfo {year} {2020})}\BibitemShut {NoStop}%
\bibitem [{\citenamefont {Adhikari}\ \emph {et~al.}(2020)\citenamefont {Adhikari}, \citenamefont {Sahoo}, \citenamefont {Mondal}, \citenamefont {Otani},\ and\ \citenamefont {Barman}}]{adhikari2020large}%
  \BibitemOpen
  \bibfield  {author} {\bibinfo {author} {\bibfnamefont {K.}~\bibnamefont {Adhikari}}, \bibinfo {author} {\bibfnamefont {S.}~\bibnamefont {Sahoo}}, \bibinfo {author} {\bibfnamefont {A.~K.}\ \bibnamefont {Mondal}}, \bibinfo {author} {\bibfnamefont {Y.}~\bibnamefont {Otani}},\ and\ \bibinfo {author} {\bibfnamefont {A.}~\bibnamefont {Barman}},\ }\href@noop {} {\bibfield  {journal} {\bibinfo  {journal} {Phys. Rev. B}\ }\textbf {\bibinfo {volume} {101}},\ \bibinfo {pages} {054406} (\bibinfo {year} {2020})}\BibitemShut {NoStop}%
\bibitem [{\citenamefont {Adhikari}\ \emph {et~al.}(2021)\citenamefont {Adhikari}, \citenamefont {Choudhury}, \citenamefont {Barman}, \citenamefont {Otani},\ and\ \citenamefont {Barman}}]{adhikari2021observation}%
  \BibitemOpen
  \bibfield  {author} {\bibinfo {author} {\bibfnamefont {K.}~\bibnamefont {Adhikari}}, \bibinfo {author} {\bibfnamefont {S.}~\bibnamefont {Choudhury}}, \bibinfo {author} {\bibfnamefont {S.}~\bibnamefont {Barman}}, \bibinfo {author} {\bibfnamefont {Y.}~\bibnamefont {Otani}},\ and\ \bibinfo {author} {\bibfnamefont {A.}~\bibnamefont {Barman}},\ }\href@noop {} {\bibfield  {journal} {\bibinfo  {journal} {Nanotechnology}\ }\textbf {\bibinfo {volume} {32}},\ \bibinfo {pages} {395706} (\bibinfo {year} {2021})}\BibitemShut {NoStop}%
\bibitem [{\citenamefont {Dai}\ and\ \citenamefont {Ma}(2021)}]{dai2021strong}%
  \BibitemOpen
  \bibfield  {author} {\bibinfo {author} {\bibfnamefont {C.}~\bibnamefont {Dai}}\ and\ \bibinfo {author} {\bibfnamefont {F.}~\bibnamefont {Ma}},\ }\href@noop {} {\bibfield  {journal} {\bibinfo  {journal} {Appl. Phys. Lett.}\ }\textbf {\bibinfo {volume} {118}} (\bibinfo {year} {2021})}\BibitemShut {NoStop}%
\bibitem [{\citenamefont {Ma}\ \emph {et~al.}(2023)\citenamefont {Ma}, \citenamefont {Li}, \citenamefont {Hao}, \citenamefont {Ong},\ and\ \citenamefont {Chai}}]{ma2023strong}%
  \BibitemOpen
  \bibfield  {author} {\bibinfo {author} {\bibfnamefont {K.}~\bibnamefont {Ma}}, \bibinfo {author} {\bibfnamefont {C.}~\bibnamefont {Li}}, \bibinfo {author} {\bibfnamefont {Z.}~\bibnamefont {Hao}}, \bibinfo {author} {\bibfnamefont {C.}~\bibnamefont {Ong}},\ and\ \bibinfo {author} {\bibfnamefont {G.}~\bibnamefont {Chai}},\ }\href@noop {} {\bibfield  {journal} {\bibinfo  {journal} {Physical Review B}\ }\textbf {\bibinfo {volume} {108}},\ \bibinfo {pages} {094422} (\bibinfo {year} {2023})}\BibitemShut {NoStop}%
\bibitem [{\citenamefont {Hayashi}\ \emph {et~al.}(2023)\citenamefont {Hayashi}, \citenamefont {Shiota}, \citenamefont {Ishibashi}, \citenamefont {Hisatomi}, \citenamefont {Moriyama},\ and\ \citenamefont {Ono}}]{hayashi2023observation}%
  \BibitemOpen
  \bibfield  {author} {\bibinfo {author} {\bibfnamefont {D.}~\bibnamefont {Hayashi}}, \bibinfo {author} {\bibfnamefont {Y.}~\bibnamefont {Shiota}}, \bibinfo {author} {\bibfnamefont {M.}~\bibnamefont {Ishibashi}}, \bibinfo {author} {\bibfnamefont {R.}~\bibnamefont {Hisatomi}}, \bibinfo {author} {\bibfnamefont {T.}~\bibnamefont {Moriyama}},\ and\ \bibinfo {author} {\bibfnamefont {T.}~\bibnamefont {Ono}},\ }\href@noop {} {\bibfield  {journal} {\bibinfo  {journal} {Appl. Phys. Express}\ }\textbf {\bibinfo {volume} {16}},\ \bibinfo {pages} {053004} (\bibinfo {year} {2023})}\BibitemShut {NoStop}%
\bibitem [{\citenamefont {Dion}\ \emph {et~al.}(2023)\citenamefont {Dion}, \citenamefont {Stenning}, \citenamefont {Vanstone}, \citenamefont {Holder}, \citenamefont {Sultana}, \citenamefont {Alatteili}, \citenamefont {Martinez}, \citenamefont {Kaffash}, \citenamefont {Kimura}, \citenamefont {Kurebayashi} \emph {et~al.}}]{dion2023ultrastrong}%
  \BibitemOpen
  \bibfield  {author} {\bibinfo {author} {\bibfnamefont {T.}~\bibnamefont {Dion}}, \bibinfo {author} {\bibfnamefont {K.~D.}\ \bibnamefont {Stenning}}, \bibinfo {author} {\bibfnamefont {A.}~\bibnamefont {Vanstone}}, \bibinfo {author} {\bibfnamefont {H.~H.}\ \bibnamefont {Holder}}, \bibinfo {author} {\bibfnamefont {R.}~\bibnamefont {Sultana}}, \bibinfo {author} {\bibfnamefont {G.}~\bibnamefont {Alatteili}}, \bibinfo {author} {\bibfnamefont {V.}~\bibnamefont {Martinez}}, \bibinfo {author} {\bibfnamefont {M.~T.}\ \bibnamefont {Kaffash}}, \bibinfo {author} {\bibfnamefont {T.}~\bibnamefont {Kimura}}, \bibinfo {author} {\bibfnamefont {H.}~\bibnamefont {Kurebayashi}}, \emph {et~al.},\ }\href@noop {} {\bibfield  {journal} {\bibinfo  {journal} {arXiv preprint arXiv:2306.16159}\ } (\bibinfo {year} {2023})}\BibitemShut {NoStop}%
\end{thebibliography}%
\end{document}

\begin{center}
{\fontsize{13pt}{16pt}\selectfont \textbf{Supplemental material\\
Exploration of magnon-magnon coupling in an antidot lattice: \\
\vspace{5pt}
The role of non-uniform magnetization texture}}\\
\vspace{14pt}
{\fontsize{10pt}{13pt}\selectfont Mathieu Moalic, Mateusz Zelent, Krzysztof Szulc, and Maciej Krawczyk}\\
{\fontsize{10pt}{11pt}\selectfont \textit Institute of Spintronics and Quantum Information,\\
Faculty of Physics, Adam Mickiewicz University, Poznan, Poland}\\
{(Dated: January 19, 2024)}
\end{center}

\begin{figure*}[h!]
    \includegraphics[width=0.86\textwidth]{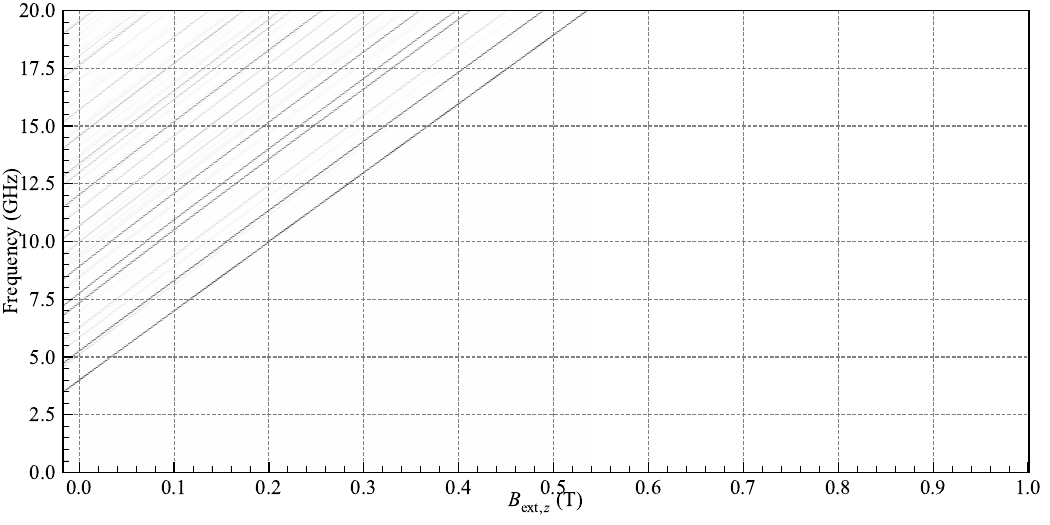}
    \caption{
    Evolution of the SW resonance spectra for the antidot lattice in dependence on the static external out-of-plane magnetic field. The line intensity corresponds to the SW amplitude. This is depicted in green in Fig.~2 of the main text.
    }
    \label{fig:suppl:ADL}
\end{figure*}
\begin{figure*}[h!]
    \includegraphics[width=0.86\textwidth]{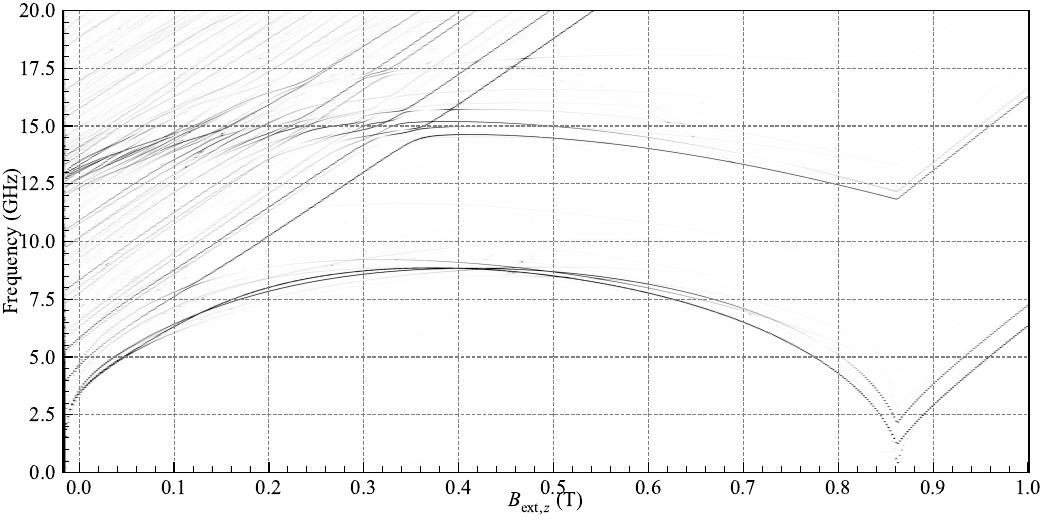}
    \caption{
    Evolution of the SW resonance spectra for the antidot lattice with modified rim in dependence on the static external out-of-plane magnetic field. The line intensity corresponds to the SW amplitude. 
    This is depicted in red in Fig.~2, 3 and 4 of the main text.
    }
    \label{fig:suppl:ADL-MR}
\end{figure*}
\begin{figure*}[h!]
    \includegraphics[width=0.99\textwidth]{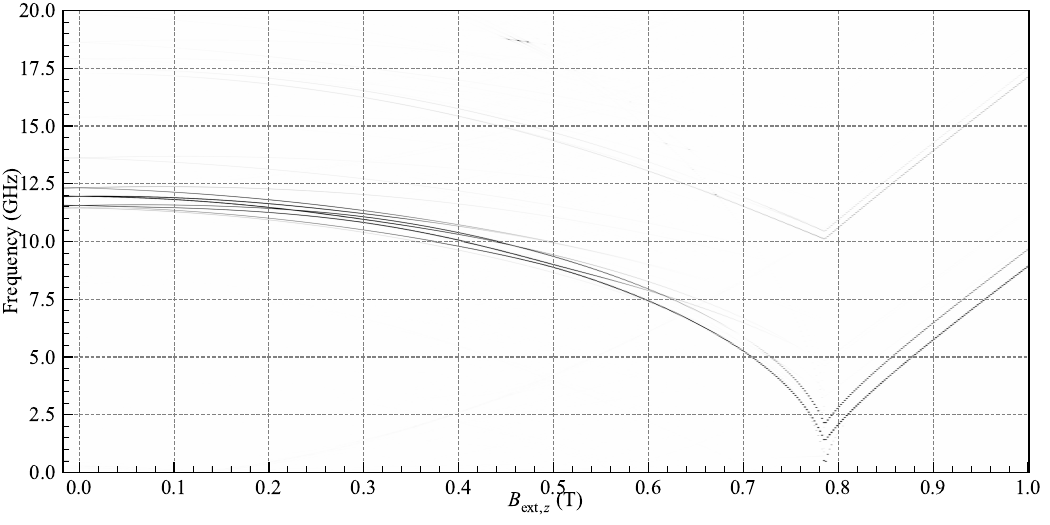}
    \caption{
    Evolution of the SW resonance spectra for the rings lattice in dependence on the external out-of-plane magnetic field. The line intensity corresponds to the SW amplitude. This is depicted in purple in Fig.~2 and 4 of the main text.
    }
    \label{fig:suppl:RL}
\end{figure*}
\begin{figure*}[h!]
    \includegraphics[width=0.99\textwidth]{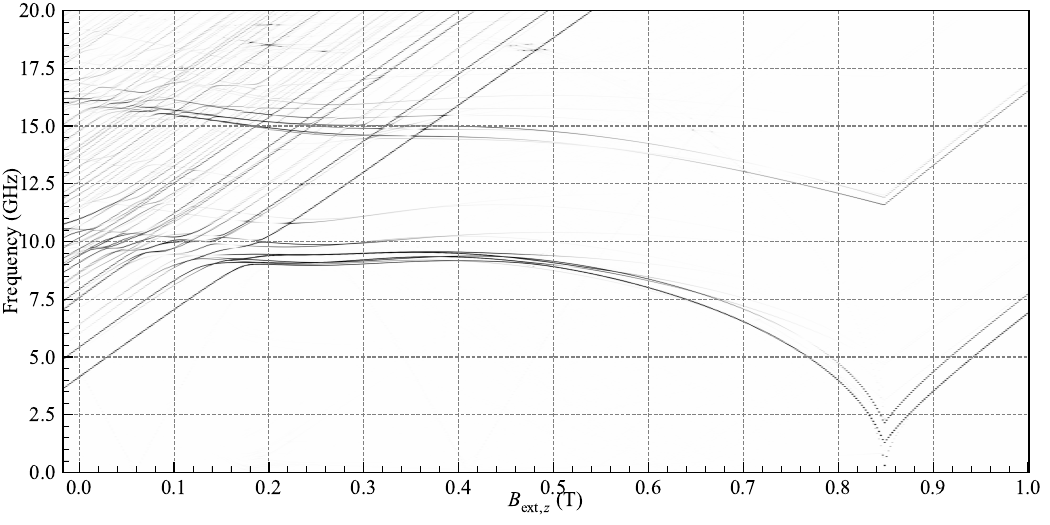}
    \caption{
    Evolution of the SW resonance spectra for the antidot lattice with modified rim with the separation between bulk and rim in dependence on the static external out-of-plane magnetic field. The line intensity corresponds to the SW amplitude. This is depicted in blue in Fig.~4 of the main text.
    }
    \label{fig:suppl:ADL-S-MR}
\end{figure*}
\begin{figure*}[h!]
    \includegraphics[width=0.99\textwidth]{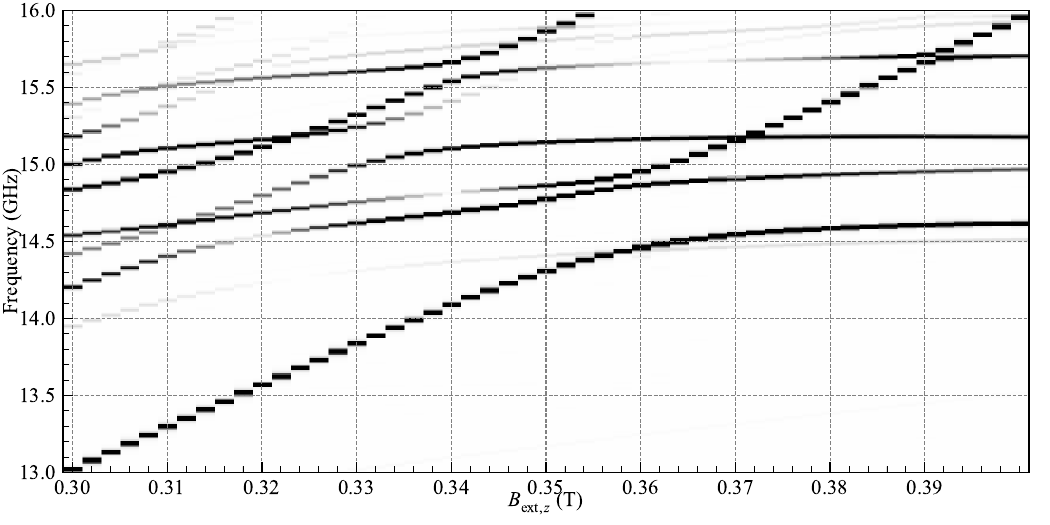}
    \caption{
    A zoom-in on Fig.~\ref{fig:suppl:ADL-MR} which represents the evolution of the SW resonance spectra for the antidot lattice with modified rim in dependence on the static external out-of-plane magnetic field. The line intensity corresponds to the SW amplitude. This is depicted in red in Fig.~5 of the main text.
    }
    \label{fig:suppl:zoom-ADL-MR}
\end{figure*}
\begin{figure*}[h!]
    \includegraphics[width=0.99\textwidth]{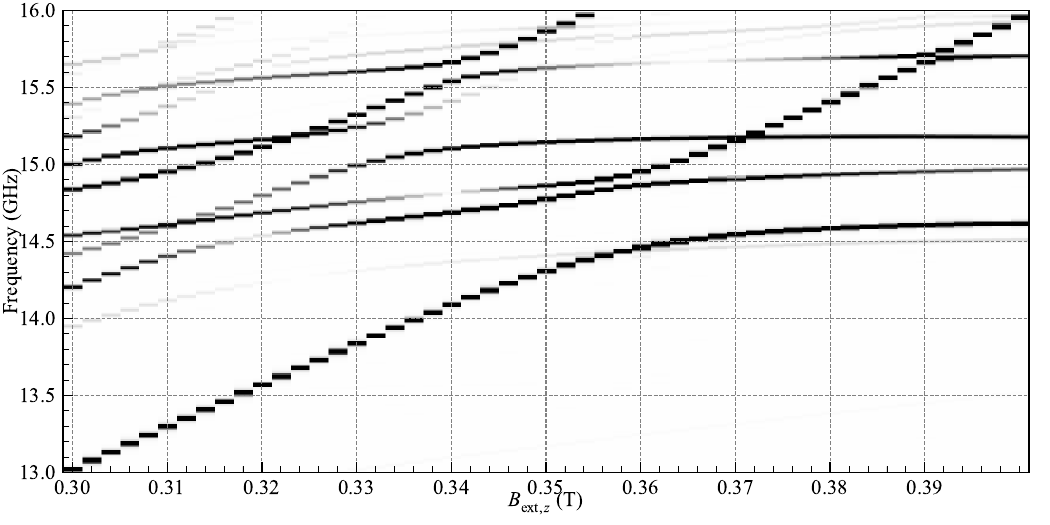}
    \caption{
    A zoom-in on Fig.~\ref{fig:suppl:ADL-S-MR} which represents the evolution of the SW resonance spectra for the antidot lattice with modified rim with the separation between bulk and rim in dependence on the external out-of-plane magnetic field. The line intensity corresponds to the SW amplitude. This is depicted in blue in Fig.~5 of the main text.
    }
    \label{fig:suppl:zoom-ADL-S-MR}
\end{figure*}
\begin{figure*}[h!]
    \includegraphics[width=0.99\textwidth]{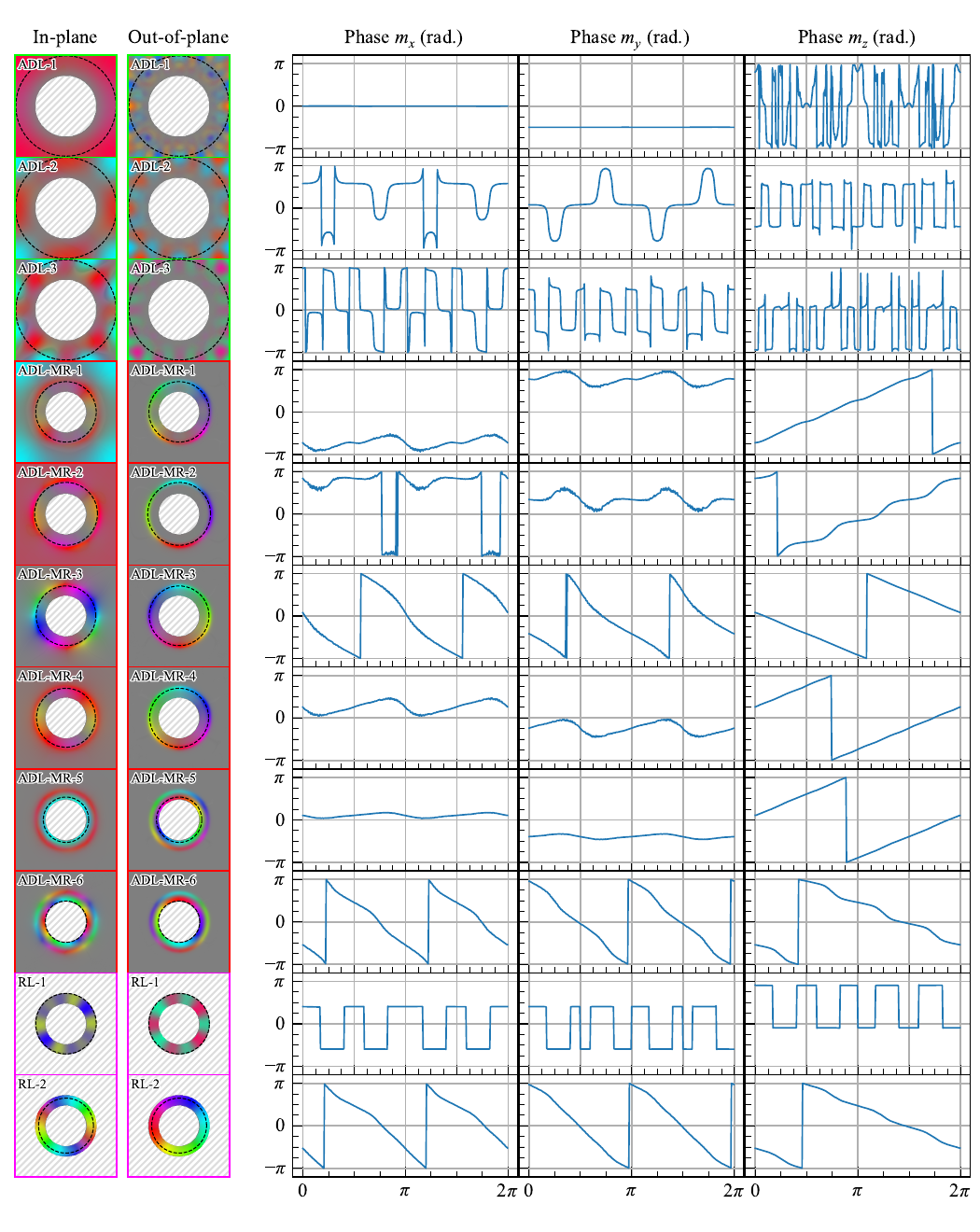}
    \caption{
    Detailed views of each mode from Fig.~2 in the main text are presented. The first and second columns represent the in-plane and out-of-plane dynamic components of the magnetization for the selected modes, respectively. The hue indicates the phase of the in-plane dynamic magnetization, while the saturation reflects its spatial amplitude. The dotted line highlights the location of the peak amplitude of each selected mode. The third, fourth, and fifth columns display the cross-sections of the phase along the dotted line, which circles the rim of the mode, for the $m_x$, $m_y$ and $m_z$ components, respectively. The azimuthal order of the modes and their chirality are deduced from the phase of the $m_z$ component.
    }
    \label{fig:suppl:modes1}
\end{figure*}
\begin{figure*}[h!]
    \includegraphics[width=0.99\textwidth]{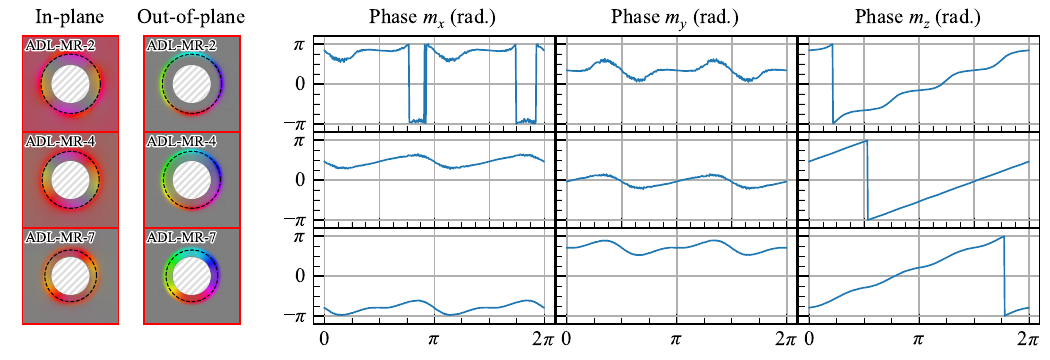}
    \caption{
    Detailed views of each mode from Fig.~3 in the main text are presented. The first and second columns represent the in-plane and out-of-plane dynamic components of the magnetization for the selected modes, respectively. The hue indicates the phase of the in-plane dynamic magnetization, while the saturation reflects its spatial amplitude. The dotted line highlights the location of the peak amplitude of each selected mode. The third, fourth, and fifth columns display the cross-sections of the phase along the dotted line, which circles the rim of the mode, for the $m_x$, $m_y$ and $m_z$ components, respectively. The azimuthal order of the modes and their chirality are deduced from the phase of the $m_z$ component.
    }
    \label{fig:suppl:modes2}
\end{figure*}
\begin{figure*}[h!]
    \includegraphics[width=0.99\textwidth]{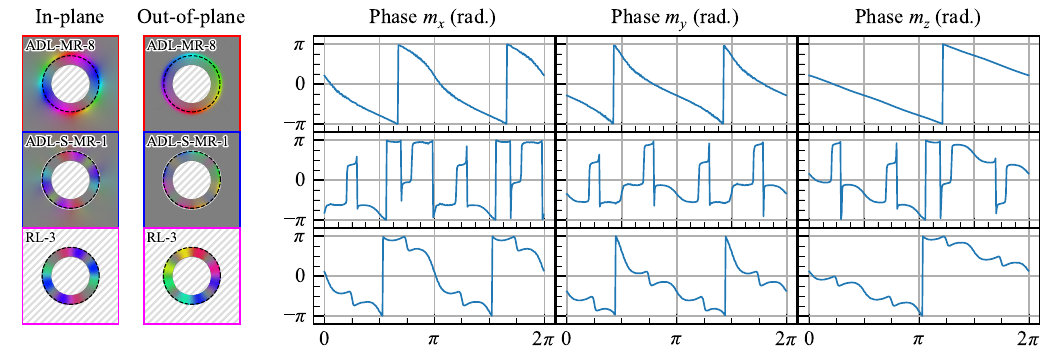}
    \caption{
    Detailed views of each mode from Fig.~4 in the main text are presented. The first and second columns represent the in-plane and out-of-plane dynamic components of the magnetization for the selected modes, respectively. The hue indicates the phase of the in-plane dynamic magnetization, while the saturation reflects its spatial amplitude. The dotted line highlights the location of the peak amplitude of each selected mode. The third, fourth, and fifth columns display the cross-sections of the phase along the dotted line, which circles the rim of the mode, for the $m_x$, $m_y$ and $m_z$ components, respectively. The azimuthal order of the modes and their chirality are deduced from the phase of the $m_z$ component.
    }
    \label{fig:suppl:modes3}
\end{figure*}
\begin{figure*}[h!]
    \includegraphics[width=0.99\textwidth]{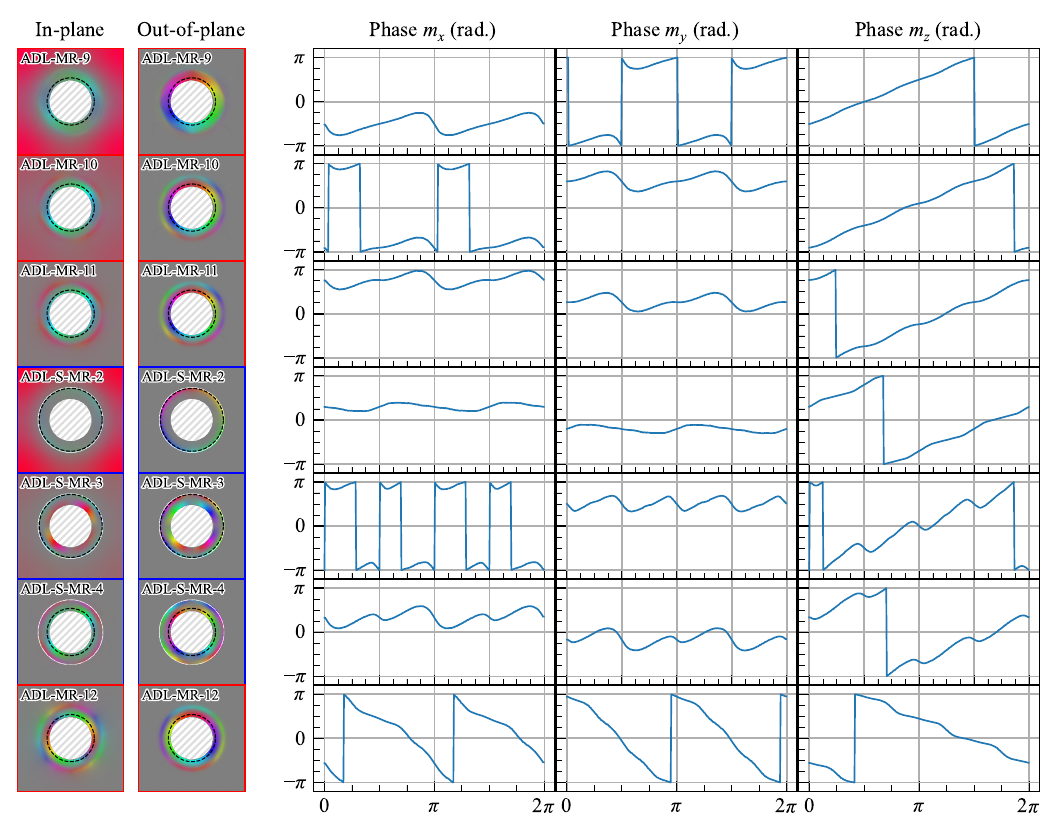}
    \caption{
    Detailed views of each mode from Fig.~5 in the main text are presented. The first and second columns represent the in-plane and out-of-plane dynamic components of the magnetization for the selected modes, respectively. The hue indicates the phase of the in-plane dynamic magnetization, while the saturation reflects its spatial amplitude. The dotted line highlights the location of the peak amplitude of each selected mode. The third, fourth, and fifth columns display the cross-sections of the phase along the dotted line, which circles the rim of the mode, for the $m_x$, $m_y$ and $m_z$ components, respectively. The azimuthal order of the modes and their chirality are deduced from the phase of the $m_z$ component.
    }
    \label{fig:suppl:modes4}
\end{figure*}